\documentclass [12pt,twoside]{article}           
\usepackage[left,modulo]{lineno}
\usepackage{setspace}

\countdef\arxiv=201

\ifnum\arxiv=0 \input cpreb.tex \fi

\ifnum\arxiv=1

\usepackage{epsfig,times,lscape}
\usepackage[usenames]{color}

    \ifnum\count102=1

    \fi

    \ifnum\count102=2
\fi

\newcommand{\nc}{\newcommand}

     \ifnum\count101=1
\nc{\qI}[1]{\section{{#1}}}
\nc{\qA}[1]{\subsection{{#1}}}
\nc{\qun}[1]{\subsubsection{{#1}}}
\nc{\qa}[1]{\paragraph{{#1}}}

\def\qpar{\vskip 2mm plus 0.2mm minus 0.2mm}
\def\qL{\hfill \break}
     \fi 

      \ifnum\count101=2
 \nc{\qI}[1]{\parindent=0mm \vskip 8mm 
{\centerline{\LARGE \color{red}#1}}\vskip 3mm}
\nc{\qA}[1]{\vskip 2.5mm \noindent 
{{\bf\large\color{blue}  #1}} \vskip 1mm \parindent=0mm}
 \nc{\qun}[1]{\vskip 1mm \noindent {\sl #1 }\quad }

\def\qL{\hfill \break}
\def\qpar{\vskip 2mm plus 0.2mm minus 0.2mm}

      \fi

\def\qth{\vrule height 12pt depth 0pt width 0pt}
\def\qtb{\vrule height 0pt depth 5pt width 0pt}

\nc{\qfoot}[1]{\footnote{{#1}}}

      \ifnum\count101=1
\def\qbu{\hfill \par \hskip 6mm $ \bullet $ \hskip 2mm}
\def\qee#1{\hfill \par \hskip 6mm (#1) \hskip 2 mm}
      \fi
      \ifnum\count101=2
\def\qbu{\hfill \par \hskip 4mm $ \bullet $ \hskip 2mm}
\def\qee#1{\hfill \par \hskip 4mm (#1) \hskip 2 mm}
      \fi

\def\qparr{ \vskip 1.0mm plus 0.2mm minus 0.2mm \hangindent=10mm
\hangafter=1}

     \ifnum\count101=1 
 \def\qdec#1{\parindent=0mm\par {\leftskip=2cm {#1} \par}}
     \fi
     \ifnum\count101=2
  \def\qdec#1{\parindent=0mm \par {\leftskip=1cm {#1} \par}}
  
  \def\qcitb#1{\noindent \hbox to 102mm{\hfill \small #1} \vskip 1mm}
      \fi

 \def\qpages#1{\count102=0{\loop\advance\count102 by 1
 \null \vfill\eject \ifnum\count102<#1 \repeat}}

\def\qn#1{\eqno \hbox{(#1)}}

\def\qth{\vrule height 12pt depth 0pt width 0pt}
\def\qtb{\vrule height 0pt depth 5pt width 0pt}

\def\qv{\vskip 0.1mm plus 0.05mm minus 0.05mm}
\def\qhu{\hskip 0.6mm}
\def\qhv{\hskip 3mm}

\def\qhw{\hskip 1.5mm}
\def\qleg#1#2#3{\noindent {\bf \small #1\qhw}{\small #2\qhw}{\it \small #3}\qv }
\fi

\begin{document}
\thispagestyle{empty}



\markboth{{\sl \hfill  \hfill \protect\phantom{3}}}
        {{\protect\phantom{3}\sl \hfill  \hfill}}

\color{yellow} 
\hrule height 20mm depth 10mm width 170mm 
\color{black}
\vskip -1.8cm 

 \centerline{\bf \Large The wage transition in developed countries and}
\vskip 2mm
 \centerline{\bf \Large its implications for China}
\vskip 20mm
\centerline{\large 
Belal Baaquie$ ^1 $,
Bertrand M. Roehner$ ^2 $, 
Qinghai Wang$ ^3 $
}

\vskip 8mm
\large

{\bf Abstract}\quad
The expression ``wage transition'' refers to the fact that
over the past two or three decades 
in all developed economies wage increases have leveled off.
There has been a widening divergence and decoupling between
wages on the one hand and GDP per capita
on the other hand. Yet, in China 
wages and GDP per capita climbed in sync (at least up to now).\qL
In the first part of the paper we present comparative statistical
evidence which measures the extent of the wage 
transition effect.\qL
In a second part we consider the reasons of this
phenomenon, in particular we explain how
the transfers of labor from low productivity sectors
(such as agriculture) to high productivity sectors (such as
manufacturing) are the driver of productivity
growth, particularly through their synergetic effects.
Although rural flight represents only one of these
effects, it is certainly the most visible because of the
geographical relocation that it implies; it is also 
the most well-defined statistically. Moreover,
it will be seen that it is
a good indicator of the overall productivity and attractivity of 
the non-agricultural sector.\qL
Because this model accounts fairly well
for the observed evolution in industrialized countries,
we use it to predict the rate of Chinese economic growth
in the coming decades. Our forecast for the average annual  
growth of real wages ranges from 4\% to 6\% depending on
how well China will control the development of its healthcare
industry.

\vskip 10mm
\centerline{\it Provisional. Version of 29 April 2016. 
Comments are welcome.}
\vskip 10mm

{\small Key-words: productivity, demographic transition, 
economic growth,
wage, earnings, agriculture, industry, primary sector,
secondary sector, tertiary sector.
\vskip 4mm

{\normalsize
1: Physics Department, National University of Singapore.
Email: belalbaaquie@gmail.com\qL
2: Institute for Theoretical and High Energy Physics (LPTHE),
University Pierre and Marie Curie, Paris, France. 
Email: roehner@lpthe.jussieu.fr\qL
3: Physics Department, National University of Singapore.
Email: qhwang@nus.edu.sg
}

\vfill\eject

\large

\qI{Introduction}

\qA{Post-industrial societies}

Nowadays (i.e. in early 2016)  the dominant conception is that the
post-industrial stage reached by most developed countries
and in which at least
75\% of employment is in the service sector is a superior
stage. It seems to realize the dream of clean, energy efficient
and highly productive economies. Clean and energy efficient they may be,
but are post-industrial societies also highly productive?
This is one of the questions that we try to answer in the present
paper.
\qpar

Perhaps it may be helpful to explain what lead us to question
the mainstream conception. Doubts were raised step-wise by
a number of observations among which one can highlight the following.
\qbu In the past two or three decades, following in the footsteps of
New York and other American cities,  home-delivery of pizzas
started in many west European cities. This marked a sharp
break with the period following World War II. In this period
labor had been considered as a rare and costly resource 
which had to be economized as much as possible. 
Home-delivery of pizzas was a low productivity
business which could be profitable only if served by poorly
paid employees. It was the harbinger of the creation of
a whole range of jobs similarly characterized by poor pay
and low productivity. As emphasized by economists such
as Jean Fourasti\'e, the fact that the
price of hair cuts or opera tickets remained unchanged 
in the course of time strongly
suggests that the tertiary sector holds little promises
for productivity improvement. At the end of the paper
we will propose a broader explanation which relies on the
fact that most service goods have little or no 
synergetic potential.
\qbu The second shock was the observation that
in the United States the average hourly wage had peaked around
1975 and stagnated ever since%
\qfoot{As a matter of fact there was a sudden change in the
trend of many other economic (e.g. income inequality)
and social indicators (e.g. proportion of inmate population).
These changes are documented more fully in Roehner (2009,
chapters 9-11).}
.
True, the Gross Domestic
Product (GDP) per capita had continued to increase but
this was due to concomitant changes, for instance the
fact that an increasing number of house wives took a full
time job or the fact that an increasing share of 
income comes from non-salary earnings.
\qbu The third element which triggered our questioning
came from the observation of service-centered economies.
Switzerland and Singapore embody fairly well the dream
of an efficient, environment-friendly
post industrial economy%
\qfoot{Although very different in terms of territory and area,
the two countries
are in the same class in terms of population (8 and 5,4 million
respectively) and GDP per capita (US\$ 84,000 and 55,000); the data
are for 2013.}%
. 
In addition both countries similarly benefited from 
massive capital inflows and foreign direct investments.
Yet, in spite of such favorable conditions
they show the same syndrome of levelling off wages
as do other developed countries.
For Switzerland this can be seen in Fig. 3 and
for Singapore it can be observed that whereas wages 
have been multiplied
by 5.5 in the 4 decades from 1965 to 2005, they have stagnated
over the last decade from 2005 to 2015 (Shanmugaratnam 2015).
In orther words, even under the
best circumstances the post-industrial, service-based economy
is not up to its promises.

\count101=0  \ifnum\count101=1
If indeed  in terms of productivity
service-based economies lag behind, then this effect should
be most obvious in city-states which have little industry. 
Is that the case?\qL
At first sight it does not seem so. In Singapore, the
real (i.e. adjusted for inflation) average monthly wage
increased by 180\% from 1990 to 2015%
\qfoot{The data are from the ``Trading Economics'' website.}%
,
indeed a remarkable achievement;
This progression was not even slowed by the
 crisis of 2008-2009.
.
However, this picture must be corrected by two observations.
First of all, despite being a city-state, Singapore 
is not purely a service economy. In 2014 its manufacturing sector
represented 18\% of its GDP (down from 25\% in 2000).
In addition, its semi-industrial port activity represented 33\% of its
GDP. Thus, altogether the industrial and semi-industrial
activities represented 51\% of GDP, i.e. much more than the share
of $ 15\% $ represented by
manufacturing, transportation and warehousing in the United States%
\qfoot{Manufacturing represented $ 11.5\% $, transportation and
  warehousing $ 3.5 \% $. The data are for 2007 and the source is:
Wikipedia: sectors of US economy as percent of
GDP (1947-2009).}%
.
\qL
Secondly, it should be observed that
the Statistical office of Singapore publishes only
wage data for the whole labor force. This contrasts with
most western countries which
also publish wage data for workers, i.e. non-managerial
employees (see Appendix A in this respect).
However, through a declaration made by the Deputy Prime Minister
(Shanmugaratnam 2015) one learns that real wages for the lowest paid
workers have been multiplied by 5.5 between
1965 and 2005 but have stagnated in the last decade.\qL
In  short, even in a successful economy like Singapore
the wages of workers leveled off in the last 10 years.
\fi

\qA{Parallel with the demographic transition}

Before describing the wage transition it may be useful
to present a transition of same kind, namely the
well-known ``demographic transition''. The two transitions are of
same kind in the sense that they occur over a similar time span
and in almost all countries.
In spite of the fact that the causes of the demographic transition
are not yet well understood, this paradigm gives us a predictive
perspective which, although rough, is quite useful nevertheless.
\qpar

There are many factors which may affect women fertility, that is
to say the average number of children that a woman will have
during her life-time. Just as illustrations on can 
mention the following: 
the mean age at marriage, the infant mortality rate,
the attitude of the prevailing religion
with respect to birth control, the laws ruling inheritance, 
the place of residence (whether rural or urban%
\qfoot{The role of this factor is demonstrated in a fairly dramatic way
by the fact that the cities of Hong Kong, Macao and Singapore
have fertility rates as low as 1.17, 0.93 and 0.80 respectively
(the data are for 2014).}%
),
whether or not the family needs a second salary, the extent to which
girls have access to education. Many other factors could
be added to this list. As a matter of fact the list 
is boundless in the sense
that for each ``primary'' factor  one can cite many ``secondary''
factors. For instance, the mean age at marriage can be 
seen as a primary factor but it is affected
by countless social and economic circumstances.

%
\begin{figure}[htb]
\centerline{\psfig{width=17cm,figure=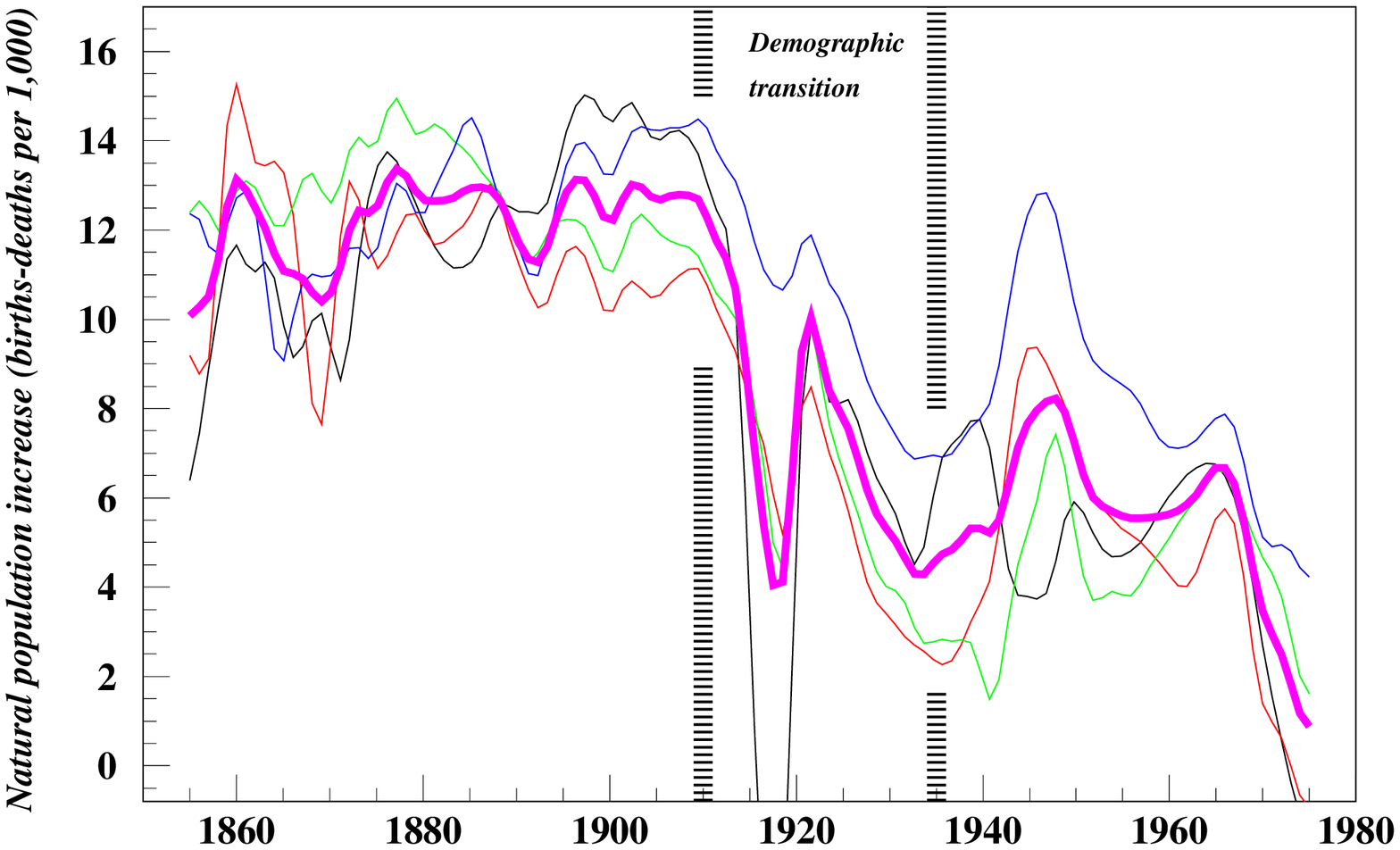}}
\qleg{Fig.\qhu 1\qhv Demographic transition in western Europe.}
{The thin lines are centered moving averages (over intervals of
5 years) of the differences between birth and death rates for
4 countries: 
Germany (black),
Denmark (blue), 
England and Wales (green).
Sweden (red). 
The thick line
is the average of the 4 cases. The demographic transition
consisted in an almost simultaneous fall of death rates (mostly
due to a decrease in infant mortality) and birth rates; however,
because the second effect was predominant the net effect
was a decrease in the rate of natural increase.
The graph shows that this transition was
marked by several ups and
downs. In addition, for some countries (e.g. the Netherlands or France)
this rule did not apply at all.}
{Source: Flora et al. (1987, vol. 2, chapter 6)}
\end{figure}
\qpar
One may think that by focusing on a single country
the question will become more clearly defined
and hence easier to solve. This is an illusion, however.
As an illustration,
consider the case of France. Between 1850 and 1940, the average fertility
rate of French women and the resulting rate of natural increase of the
French population were
much lower than in other Western European countries%
\qfoot{Detailed comparative data can be found in Flora et al. (1987)
but a broad view can be obtained just by comparing the number of
immigrants to the United States or to Latin America. There is
a strong contrast  between the massive flows of Germans and Italians
and the very small numbers of French immigrants}%
.
The explanation most commonly accepted by historians relies on the
observation that according to French inheritance laws 
all children were entitled to an equal part of their father's landholding.
Thus, it is said, to avoid splitting their land, farmers did not
wish several children. However, can such
an explanation be sufficient 
when one realizes that in 1906 only 43\%
of the labor force was working 
in agriculture and that only one half of
this percentage, that is to say some 20\% of the labor force,
were landowners (Flora et al. 1987, p. 500)?
\qpar

On the contrary, if, instead of focusing on a single case, the
scope of the analysis is broadened to a whole range of
similar countries, then a fairly well defined pattern
emerges which is called the {\it demographic transition}.

It consists in the fact that between 1900 and 1940, 
in almost all west European countries the 
annual rate of natural increase (per 10,000 population) 
fell from about 120 to less than 50%
\qfoot{For instance, in Germany the fall was from  141 to 54,
in Britain from 116 to 30, in Sweden from 111 to 25, in 
Switzerland from 100 to 46 (Bunle 1954, p. 16).
There were exceptions however, the most conspicuous of which
were Italy and the Netherlands. In both cases, the rate of natural
increase remained around 100. France should not be considered
as an exception because with its initial increase rate of only 11
per 1,000, no transition could happen.}%
.
\qpar
We do not know exactly what causes
triggered the fall in births per women
and the correlative fall in natural increase. Was it better education,
urbanization, higher income, or some other factors? 
Yet, despite possible exceptions, the observation of the 
West European demographic transition gave us a predictive rule.
This rule displayed its full usefulness when 
between 1950 and 2000 a similar
demographic transition unfolded 
in many developing countries, e.g. Brazil,
India, Indonesia, Mexico, South Korea, Turkey, Vietnam.
This transition was even of greater magnitude than the 
first in the sense that within 50 years
fertility rates fell from a high level of about 5 to
a level of 2 which is the long-term replacement rate.
As in the first transition, in this second transition there were also
a few countries which did not follow the rule (e.g. the Philippines
or Nigeria). However, it can hardly be denied that
the demographic transition pattern gives us a better understanding
and allows us to predict future trends.

\qA{The wage transition rule}

In the present paper, we wish to describe what can be called
a {\it wage transition}. It may be summarized by the following
statement:
\qdec{\it {\color{blue} Wage transition rule.}\quad 
In developed economies, after an initial phase during
which real wages increased exponentially (with a doubling-time of
about 30 years), a second phase set in around 1980 during which average
wages leveled off or even fell slightly. 
Most often this second phase set in after the share of
agriculture in the labor force had fallen under 10\%.}
\qpar

Before stating this rule
we discussed briefly the demographic
transition because the two transitions share 
several important aspects.
\qbu Although (as will be shown below)
the wage transition is confirmed by a large amount of statistical evidence,
there are a few exceptions (i.e. the UK).
\qbu The demographic transition could not
be attributed to a single factor but was rather correlated
with a whole bundle of variables; similarly the
wage transition comes along with a whole range of changes.
\qbu Despite these limitations, the wage transition rule
provides a framework that gives a better understanding
and allows testable predictions. 
\qpar

As an example of a better understanding one can mention the
case of the United States.
When this example is considered alone it might appear
surprising that wages increased sharply until
1975 and slumped thereafter up to present day. 
The observation
becomes less puzzling when seen in the light of 
the wage transition rule because then it just appears
as an early instance of a transition that took place
in most countries. This tells us that instead of trying
to explain it by purely American causes (e.g. the
Vietnam War or the decline of the stock market) 
we should rather consider
factors that are common to most industrialized countries.

\qA{Growth of the Chinese economy}

As an example of a testable prediction, one can mention
the issue of the growth of the Chinese
economy, a question much debated currently (November 2015). 
The wage transition rule and the observation that the share
of Chinese agricultural employment will not fall under 10\% until
at least 2025 suggest the following prediction. 
\qdec{After increasing exponentially (with a doubling time of 8.4 years)
from 1990 to 2015, the average wage in China should
follow the same trend for at least another decade 
that is to say until 2025.}
\qpar
During the past three decades the increase of the average wage
paralleled the growth of the Gross Domestic Product (GDP).
From 1990 to 2015 the GDP per capita (at current prices)
grew from 1,866 RMB to 49,300 RMB. With an increase
of the GDP deflator from 12 to 42 this gives a multiplication
by 7.5 for the real GDP per capita. The average annual growth rate
was 8.2\%  which corresponds to a doubling time of 8.4 years.
In a country as large as China, domestic consumption plays
a crucial and increasing role which means that
a rising average wage is both the effect and engine
of economic growth. 
\qpar

The following qualifications may apply as second order effects.
\qbu If productivity progress slows down
because of a rapid development of the healthcare sector
(there is a longer discussion of this point below)
then one would expect a somewhat slower growth, say around 6\%.
\qbu In addition, in
the coming decades it is likely that,
as in western countries, the share of non-salary earnings will
increase. As a result, salaries may grow slower
than the GDP/capita. This would lead to an annual growth forecast 
for salaries in the range $ 4\%-5\% $.
\qpar

Actually, in the previous prediction 
the key-point is not the exact value
of the rate but rather the statement that such a long-term
trend will not be derailed by one or two short-lived
recessions%
\qfoot{Just for the purpose of comparison, it can be of
interest to consider the case of Mexico.
Between 1980 and 2013 it experienced 3 major crises. 
\qbu In December 1982 the country 
was bailed out after defaulting
on its sovereign debt. It was not until 1994 that the GDP/capita
resumed its pre-crisis level. 
\qbu Then,
in early 1995 a second bail out
(\$50 billion) was necessary to prevent a new default.
\qbu  Finally, the worldwide crisis of 2008 led to a 5\% fall of GDP.
\qL
As nowadays for Greece, there were a succession of debt restructuring
plans: e.g. the Baker, Brady and Clinton plans.
Because of these recurrent crises, the GDP/capita expressed in
US dollars increased by only 31\% in the 33 years from 1980
to 2013. Expressed in constant dollars
the growth would be about zero. Thus, it
can be said that the ``North American Free Trade
Agreement'' (NAFTA) which started in 1994
has had but a dismal effect on the economic growth of 
Mexico.}%
.
Nevertheless, it may not hold if there is an
upheaval similar to what happened in the Soviet Union in the
years following 1991. 

\qA{Connected issues}

Before starting the discussion of the wage transition,
it may be helpful to keep in mind the following observation.
In economics all phenomena are inter-connected.
Therefore, it would be easy to list many effects
which may possibly play a role in the levelling off
of wages. For instance, the financialisation of the economy
contributes to the divergence between 
GDP/capita and wages; growing inequality contributes
to the divergence between high and low earnings;
through expansion of the supply side, immigration
contributes to eroding wages; the gradual disappearance
of unions may have the same effect. 
\qpar

Although all these effects may play a role, thanks to
our comparative perspective, we know that they do not play
the main role. Indeed, the great advantage of this approach
is that it allows us to filter out secondary factors. 
In Japan
there is little immigration, in Sweden unions are still
important forces, in Germany, financialisation of the economy
is much less developed than in the US. Nevertheless, in
all these countries the wage transition effect can be observed. 
In short, even if the previous effects exist, they are
only second-order effects in the sense given to this expression
in physics.

\qI{The wage transition in the United States}

\qA{Observation}

There are two good reasons for
starting this investigation with the case of the United States.
The first is that, as already said, it was the first country where 
this transition occurred. Secondly,
it is probably the country where the wage transition was the sharpest.
This can be seen clearly by comparing Fig. 2 and Fig. 3.

%
\begin{figure}[htb]
\centerline{\psfig{width=17cm,figure=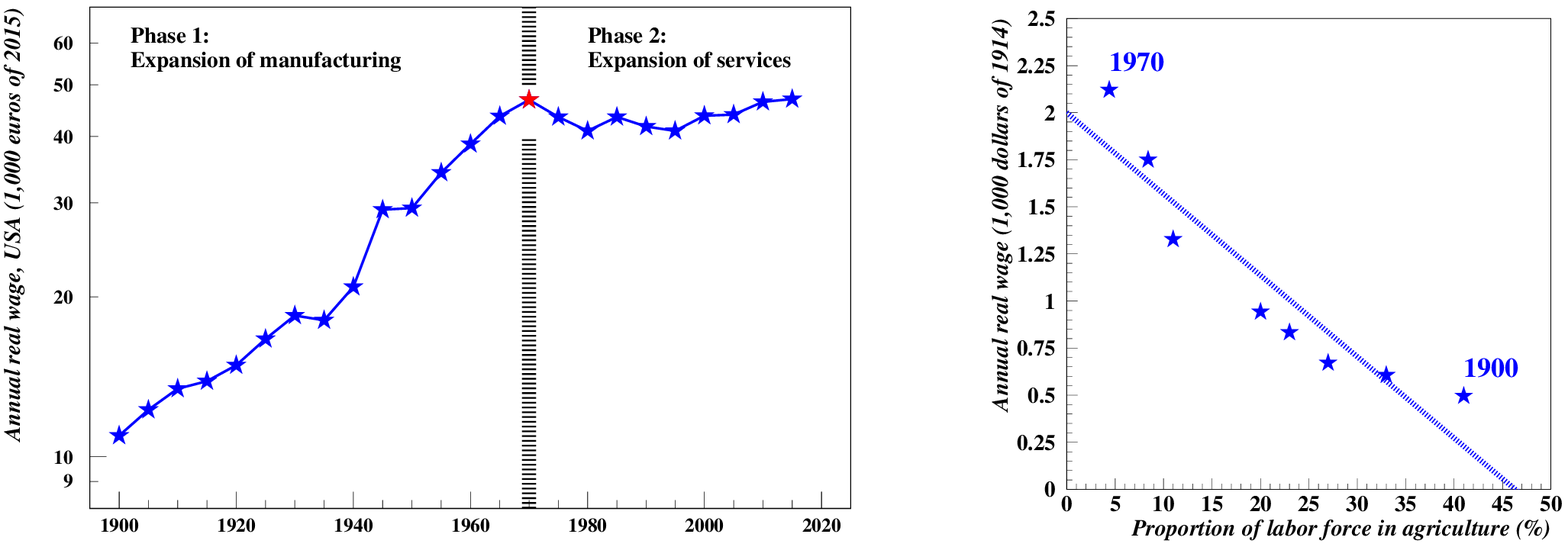}}
\qleg{Fig.\qhu 2\qhv Wage transition in the United States.}
{In phase 1 the mean wage, $ w $, increased exponentially
at an average rate of $ 2.0\% $ per year 
whereas in phase 2 it stagnated. Simultaneously
the proportion of the labor force employed in agriculture, $ f_a $
fell from $ 41\% $ in 1900 to $ 4.4\% $ in 1970. There 
is a high correlation, namely 0.987, between the two changes.
The
relationship between the logarithms of the two variables
is: $ \log w =-a\log f_a+b,\quad a=0.67 \pm 0.09,\ b=1.87\pm 0.06 $.
However, because several other economic factors changed as well 
during this time period, 
this close connection does not necessarily mean that the wage effect
resulted from the shift in the labor force.}
{Sources: Historical Statistics of the United States 1975, 
Liesner 1989.}
\end{figure}

Whereas in the US the transition from exponential increase to
stagnation occurred abruptly, elsewhere there was a smooth transition
involving a progressive reduction of the growth rate.

\qA{Possible ``explanations''}

If we limit our attention to the US case, many possible
``explanations'' may come to mind. For instance, one can mention
the following.
\qbu The decades after 1945 were marked by a progressive erosion
of the fighting capability and bargaining power
of American unions. This trend
already started with the Taft-Hartley Act of 1947 but
it really gained momentum only in the 1970s. One of its clearest
manifestations was the dramatic fall in the frequency of strikes.
More details can be found in Roehner (2009). For instance,
in the private sector the unionization rate dropped
from about 30\% in the 1950s to 8\% in 2007.
\qbu The suddenness of the transition may be ``explained'' by
a series of external shocks. 
(i) 
the confidence crisis which affected the dollar in the late 1960s
and which eventually led to President Nixon's 
action in 1971 ending its convertibility to gold. 
(ii) the slump of stock prices at the New York Stock Exchange, 
(iii) the oil shock,  
(iv) the aftershocks of the Vietnam War d\'eb\^acle.
\qpar

In order to narrow down the set of possible ``explanations''
we must consider a broader range of cases.
This is what is done in the next section. The fact that
the same wage transition can be observed in many countries will
ipso facto eliminate any specifically American explanation.

\qI{The wage transition in comparative perspective}

\qA{Average wage increases}
In February 2015, in his budget speech the Deputy Prime Minister
(and also Finance Minister) of Singapore, Tharman Shanmugaratnam,
warned that wage stagnation had set in for most
developed economies in the United States, Europe and Japan
(Shanmugaratnam 2015). As seen in the previous section
regarding the United States, and as will be seen in the present
section for the European countries and Japan, wage stagnation
is in fact a phenomenon which started over two decades ago.
However, as already mentioned,
in Singapore the phenomenon started only around 2005.
In other words, like the demographic
transition, the wage transition did not occur simultaneously
in all places but occurred gradually with 
country-dependent time lags.
This is illustrated in Fig. 3. 

%
\begin{figure}[htb]
\centerline{\psfig{width=17cm,figure=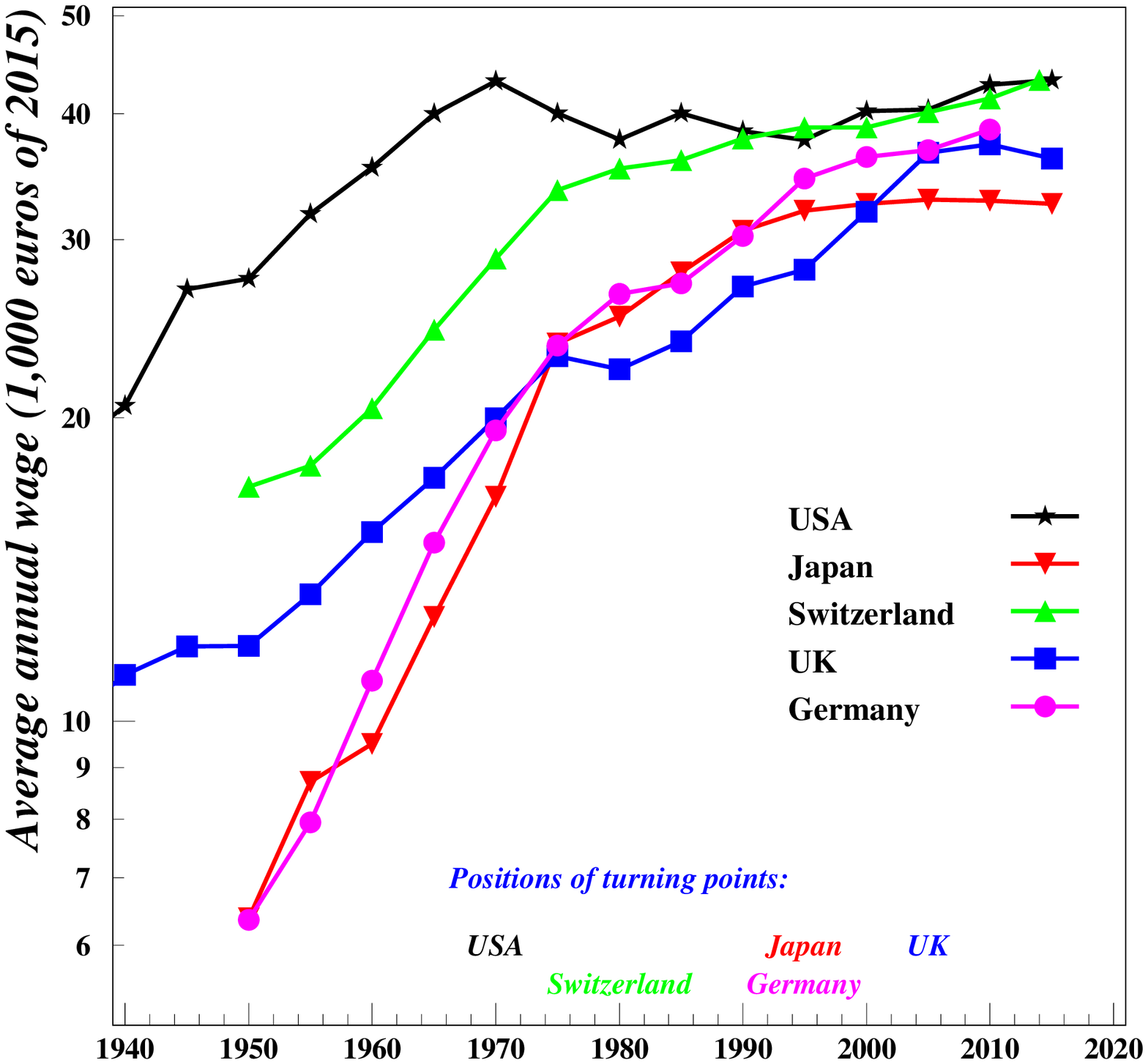}}
\qleg{Fig.\qhu 3\qhv Wage transition in several countries.}
{In phase 1 the average wage increased exponentially
whereas in phase 2 it stagnated or increased more slowly.
The names of the countries at the bottom of the graph
indicate the years when the transition occurred. To 
the 5 countries shown one could add Singapore whose transition
occurred in 2005 as discussed in the text.}
{Sources: 
USA: see Fig.1; 
Japan: Historical Statistics of Japan, average monthly contractual
earnings;
Switzerland: The website of the ``Federal Office of Statistics''
gives an annual series of real wages starting in 1939;
UK: Until 1985: Liesner 1989, 1990-2015: Office of National Statistics,
average weekly earnings of manual workers;
Germany: Until 1985: Liesner 1989, 1990-2015: Federal
Statistical Office, gross hourly earnings for manual production
workers.}
\end{figure}

\qA{The case of Britain}

It can be observed that Britain stands apart both in Fig. 3 and
in Fig. 6. One may wonder why. 
\qpar
In 1846, that is to say very early compared with other European
countries,
through the repeal of the Corn Laws which imposed heavy duties
on grain imports, Britain decided that grain supplies could
be obtained in a much cheaper way through overseas imports
particularly from the British Empire (e.g. from Canada or India%
\qfoot{To the point that India exported grains to Britain
even in years
when there were famines in some parts of the country.}%
).
This step was foreshadowed by recurrent land ``Enclosure Acts''
(e.g. those of 1801 and 1845)
which transformed into private property the
 ``common lands'' that hitherto had allowed
farming by landless farmers. 
\qpar
Through these steps  people were compelled to move from the
countryside to cities. In other European countries a substantial
rural flight started only much later.

\qA{Interpretation}

Table 1 shows that with respect to productivity
the only clear and long-term distinction
is the one between agriculture and manufacturing.
The tertiary service sector is a mixed bag involving
activities of very different productivity levels.
As a consequence the composition of this sector will 
be different from one country to another and in each
country it will change in the course of time.
For instance, the software industry which nowadays is
an important component in the United States was 
non-nonexistent some 30 years ago and is still under-represented
in European countries. Another example is the health care
industry; according to table 1, in 2012
this sector represented 12\% of the
total employment and 7.1\% of the total GDP%
\qfoot{Another indicator is the weight of health
care expenditures; according to World Bank data in 2012 
it represented 17\% of the US GDP.}%
; some 30 years ago the weight of this sector was much
smaller and nowadays it is still much lower 
in many other developed countries.
\qpar
It is this changing shape of the tertiary sector which
makes it difficult to set up a model which would have
a permanent validity. There is a similar
difficulty with the demographic transition in the sense
that the transition described by Fig. 1 was in fact
followed by a second one in the early 21th century that
brought down the fertility rates of many developed countries
well below the replacement rate of 2.1.
Although the two transitions may share some characteristics
their mechanisms may not be identical.

\begin{table}[htb]

\small

\centerline{\bf Table 1 \quad GDP per employee  in various
economic sectors (USA, 2012)}

\vskip 5mm
\hrule
\vskip 0.7mm
\hrule
\vskip 2mm

$$ \matrix{
&\hbox{Sector}\hfill &  \hbox{Contribution} & \hbox{Employment}
\hfill&
\hbox{GDP per employee} \cr
&\hbox{}\hfill &  \hbox{to GDP} & \hbox{} \hfill &\cr
\qtb
&\hbox{}\hfill &   \hbox{[billion \$]}  & \hbox{[million]}
 & \hbox{[thousand \$]}\cr
\noalign{\hrule}
\qth
1&\hbox{Educational services}\hfill & \hfill 183  & 3.34 & 54.8\cr
2&\hbox{Retail trade}\hfill & \hfill 932  & 14.9 & 62.5 \cr
3&\hbox{\color{blue} Health care, social assistance}\hfill & \hfill 1,153
&16.9 & \color{blue} 68.2 \cr
4&\hbox{Art, entertainment}\hfill & \hfill 157  & 2.09 & 75.1\cr
5&\hbox{\color{green}Agriculture, forestry, fishing }\hfill & \hfill 186 & 2.11 
 & \color{green} 88.1\cr
6&\hbox{Government (federal and local)}\hfill & \hfill 2,197  &21.9 & 100\cr
7&\hbox{Transportation, warehousing}\hfill & \hfill 467  &4.41 & 106\cr
8&\hbox{Professional services}\hfill & \hfill 1,912  &17.9 & 106\cr
&\hbox{\color{magenta} All tertiary sectors}\hfill & 
\hfill 11,929  & 97.6&  \color{magenta} 122\cr
&\hbox{\color{red} All sectors}\hfill & \hfill 14,509  & 112& 
\color{red} 129\cr
&\hbox{\color{blue} All non-agricultural sectors}\hfill & \hfill 14,323  & 110& 
\color{blue} 130\cr
9&\hbox{\color{blue} Manufacturing}\hfill & \hfill 1,983  &11.9 &
\color{blue}166\cr
10&\hbox{Wholesale trade}\hfill & \hfill 962  & 5.67 & 169\cr
11&\hbox{Information}\hfill & \hfill 737  & 2.74& 269\cr
12&\hbox{Financial activities}\hfill & \hfill 3,229  & 7.78& 415\cr
\qtb
13&\hbox{Mining}\hfill & \hfill 411  & 0.80 & 514\cr
\noalign{\hrule}
} $$
\vskip 1.5mm
\small
Notes: The GDP per employee given in the last column can be seen
as a measure of employee productivity.
Despite the modernization of US agricultural production
there is still a  1:2 productivity gap between 
agriculture and manufacturing.
This table suggests that what is usually called the service
sector is in fact a very heterogeneous category in
which productivity covers a scale from 1 to 10.
The ratio between the productivities in agriculture and 
non-agriculture sectors is equal to $ 130/88=1.47 $. 
The present table is not an exhaustive list of all sectors
but it represents 90\% of the total GDP.\qL
{\it Sources: GDP: Bureau of Economic Analysis (interactive table);
Employment: Bureau of Labor Statistics (``Table 2.1
Employment by major industry sector''.}
\vskip 5mm
\hrule
\vskip 0.7mm
\hrule
\end{table}

The global productivity of the tertiary sector is a weighted
average%
\qfoot{In fact in the present case the weighted average
is identical to the GDP per employee:
$ \sum (G_i/p_i)(p_i/P)=(1/P)\sum G_i $.}
which in the case of the sectors listed
in Table 1 (i.e. all 13 sectors except
agriculture, manufacturing and mining) is equal to \${$ 122,000 $}
per employee, a figure which is mid-way between
the productivities of agriculture and manufacturing.
\qpar
In the next section we examine the implications for
economic growth.

\qA{Implications for economic growth}

For China in 2013 the weights and productivity
of the agricultural sector on the one side and of
all non-agricultural sectors on the other side
are given in Table 2.

\begin{table}[htb]

\small

\centerline{\bf Table 2 \quad Contribution of different
sectors to the Chinese GDP, 1990-2013}

\vskip 5mm
\hrule
\vskip 0.7mm
\hrule
\vskip 2mm

$$ \matrix{
\hbox{Sector}\hfill &  \hbox{Share}  & \hbox{Share}&
\hbox{GDP per} & \hbox{Productivity}\cr
\hbox{}\hfill &  \hbox{of GDP}  & \hbox{of employment} \hfill & 
\hbox{employee}& \hbox{multiplier}\cr
\hbox{}\hfill &  g_i & e_i & (g_i/e_i)g & k=(g_i/e_i)/(g_1/e_1)\cr
\qtb
\hbox{}\hfill &   \hbox{[\% of GDP]}  & \hbox{[\% of employment]}
 & \hbox{}&\cr
\noalign{\hrule}
\qth
\hbox{\color{blue} 1990}\hfill &  &  & &\cr
\hbox{Agriculture}\hfill &  g_1=25  & e_1=60 & g_1/e_1=0.42g &1\cr
\hbox{Industry and tertiary sector}\hfill & g_2=75  & e_2=40 & 
g_2/e_2=1.87g & 4.3 \cr
\hbox{\color{blue} 2013}\hfill &    &  & \cr
\hbox{Agriculture}\hfill &  g_1=10  & e_1=34 & g_1/e_1=0.29g & 1 \cr
\qtb
\hbox{Industry and tertiary sector}\hfill &  g_2=90  & e_2=66 & g_2/e_2=1.36g &
4.5 \cr
\noalign{\hrule}
} $$
\vskip 1.5mm
\small
Notes: $ g $ denotes the GDP per employee for the whole economy.
Between 1990 and 2013 the percentage of the total labor force
engaged in the agricultural sector fell from 60\% to 34\%.
In 2013 the productivity gap between 
agriculture and the rest of the economy was 4.5; this was three times
the ratio of 1.47 found in Table 1 for the United States.
Needless to say, even in agriculture there was 
a marked productivity increase due to the rapid growth of $ g $.
\qL
{\it Sources: GDP data: Bajpai (2014); employment data: World Bank,
(Interactive table entitled ``Employment in agriculture (\% of total
employment)'').}
\vskip 5mm
\hrule
\vskip 0.7mm
\hrule
\end{table}

With respect to rural flight, 
in the United States
the major transformation took place between 1900 and 1950.
In this time span the percentage of the labor force engaged
in agriculture fell from 41\% to 11\%,
that is to say at an annual rate of 0.6\%.
Based on the data of Table 1 and 2, Fig. 4 summarizes
the respective situations of the United States and China.
In the United States between 2000 and 2010 the percentage
of the labor force employed in agriculture fell
from 2.5\% to 1.5\%, that is to say at an annual rate of 0.1\%.
%
\begin{figure}[htb]
\centerline{\psfig{width=12cm,figure=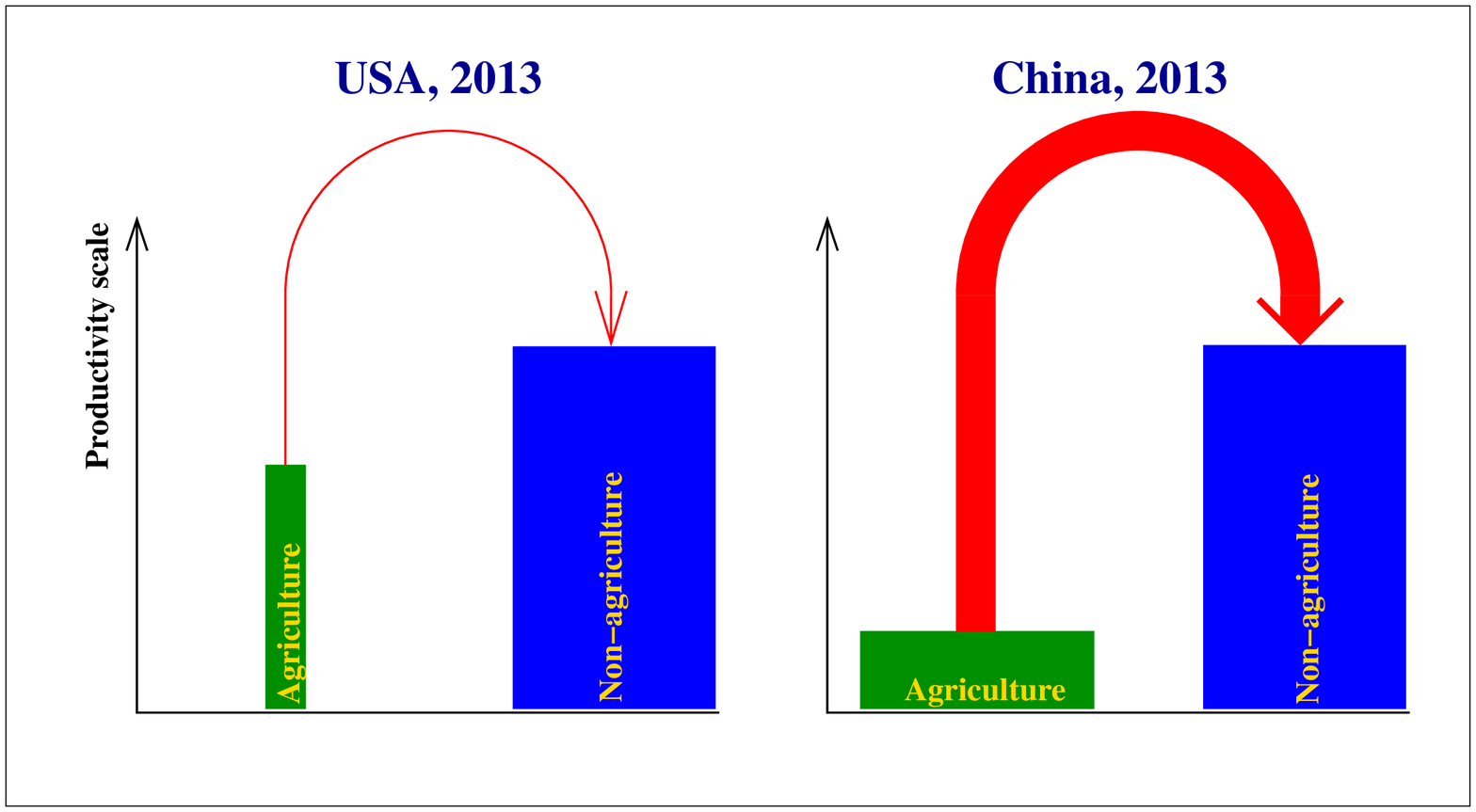}}
\qleg{Fig.\qhu 4\qhv Transfer of rural population
to the non-agricultural sector.}
{In China the transfer rate was over 10 times faster than
in the United States.The figure respects the order of magnitude
of the productivity gap within each country and also
the magnitude of the respective rates of transfer.
The scales of the two vertical productivity axes
are not the same; this allowed us to draw
two identical rectangles to represent the non-agricultural
sectors.}
{Sources: Tables 1 and 2.}
\end{figure}

In China during the same time interval, this percentage
fell from 50\% to 36\%, that is to say at an annual rate
of 1.4\%. 
Whereas nowadays in the United States
this transfer of jobs from the agricultural sector to the 
non-agricultural sector represents a very marginal
source of productivity progress, in the first half of the
20th century it was a major source of productivity growth.
However, the present transfer rate in China is twice
as fast as it was at that time in the US. It was between 1940 and
1950 that the US transfer rate reached its maximum
with an annual rate of 0.9\%, still lower than 
seen in China in the past decades.

\qI{Model describing the transfer effect}

In order to test to what extent the mechanism described in
Fig. 4  can explain long-term wage increases we need to
set up a model along these lines.

\qA{Definition of key-variables}
In order to make it as simple as possible, we will 
introduce only two sectors: agriculture (sector $ 1 $) on the one hand 
and the rest of the economy (sector $ 2 $) on the other hand.
At first sight this may seem a rough approximation;
for instance, mining and health care will be together
in sector $ 2 $; however, mining is a fairly small
sector (it is 21 times smaller than health care).
The purpose of our two-sector model is to provide
a broad picture and to
explain (and predict) long-term trends.
\qpar

We introduce the following notations:\qL
$ E $: total employment, $ E_i,\ i=1,2 $: employment in sectors 1,2
respectively. $ e_i=E_i/E $.\qL
$ G $: total (real) GDP, $ G_i,\ i=1,2 $: GDP of sectors 1,2 respectively,
$ g_i=G_i/G $.\qL
We denote by $ p=G/E $ the GDP per employee of the whole economy.
$ p_i=G_i/E_i=(g_i/e_i)(G/E)=(g_i/e_i)p $ denotes the GDP per employee
in each sector, that is to say their productivity.
\qpar

Now we introduce a variable that we call the {\it productivity 
multiplier} that will characterize how much the productivity
of sector 2 is higher than the one of sector 1.
 $$ \hbox{Productivity multiplier:}\quad k={ p_2 \over p_1 }=
{ g_2/e_2 \over g_1/e_1 } = { (1-g_1)/(1-e_1)\over g_1/e_1 } $$

The interesting point is that $ k $ can be computed as soon as 
one knows the two numbers $ g_1,e_1 $.

\qA{Order of magnitude and changes of key-variables}

In order to give some substance to these definitions, let us
give some numerical values for the United States and China.
The data for $ p $ are expressed in dollars of 1982 and renminbi of 2013
respectively.
\qbu {\color{blue} USA}: \qL
{\small
$ 1900:\ p=\$9,593,\ e_1=41\%,\ g_1=23\%,\ g_1/e_1=0.56,\ k=2.32 $\qL
$ 1950:\ p=\$20,424,\ e_1=11\%,\ g_1=6.9\%,\ g_1/e_1=0.62,\ k=1.66$\qL
$ 1980:\ p=\$32,095,\ e_1=3.4\%,\ g_1=3.0\%,\ g_1/e_1=0.88,\ k=1.14$\qL
}
\qbu {\color{blue} China}: \qL
{\small
$ 1983:\ p=\hbox{RMB}\ 5,114,\ e_1=67\%,\ g_1=33\%,\ g_1/e_1=0.49,\ k=4.12$\qL
$ 2000:\ p=\hbox{RMB}\ 24,184,\ e_1=50\%,\ g_1=50\%,\ g_1/e_1=0.30,\ k=5.67$\qL
$ 2013:\ p=\hbox{RMB}\ 73,513,\ e_1=31\%,\ g_1=10\%,\ g_1/e_1=0.32,\ k=4.04$\qL
}

A noteworthy point is the stability of $ g_1/e_1 $; it
changes much less than $ e_1 $ and $ g_1 $ separately and also much 
less than $ p $. How should this be interpreted?
The fact that $ g_1/e_1=p_1/p $ is quasi-constant in spite of the
fact that $ p $ doubles, triples or quadruples 
means that agricultural productivity
increases approximately at the same pace as global productivity.
Moreover, it can be observed that $ k $ changes fairly slowly
which for a structural variable is of course a welcome characteristic.
\qpar

Other data will be found in Table 3.
In addition, intermediate values of the GDP per employee
for 1995, 2000, 2005 and  2010 are as follows (expressed in 2015 RMB):\qL
$ p_1: 7,434,\ 7,547,\ 10,150,\ 20,030 $ \qL
$ p_2: 27,323,\ 42,189,\ 60,273,\ 75,394 $\qL
During these 25 years the GDP per employee
has been increasing exponentially both in the agricultural sector
and in the rest of the economy. The coefficients of correlation
$ (t,\log p_i) $ are 0.97 and 0.99 respectively. The exponents
are given in the notes of Table 3.

\begin{table}[htb]

\small

\centerline{\bf Table 3:\quad China: GDP per employee in agriculture
and rest of economy, 1990-2015}

\vskip 5mm
\hrule
\vskip 0.7mm
\hrule
\vskip 2mm

$$ \matrix{
 & \hbox{} &  & 1990 & 2015 & \hbox{Ratio} \cr
\qtb
 & \hbox{} &  &  &  & 2015/1990 \cr
\noalign{\hrule}
\qth
 & \hbox{\bf \color{blue} Whole economy} \hfill& &  &  &
\hbox{} \cr
1 & \hbox{GDP (billion of current RMB)} \hfill&  & 1,866  & 67,000 &
\hfill 36.00 \cr
2 & \hbox{GDP deflator} \hfill&  & 12  & 42 & \hfill 3.50 \cr
3 & \hbox{GDP (billion of 2015 RMB)} \hfill& G & 6,531  & 67,000 &
\hfill 10.30 \cr
4 & \hbox{Labor force (billion)} \hfill& E & 0.64  & 0.81 &  \hfill 1.26\cr
5 & \hbox{GDP per employee (2015 RMB)} \hfill& p & 10,204  & 82,716 &
\hfill 8.10\cr
 & \hbox{\bf \color{blue} Agriculture} \hfill&  & &  &  \cr
6 & \hbox{GDP (billion of 2015 RMB)} \hfill& G_1 & 1,763  & 6,030  &
\hfill 3.42\cr
7& \hbox{Labor force (billion)} \hfill& E_1 & 0.33  & 0.25 & \hfill 0.75 \cr
8& \hbox{GDP per employee (2015 RMB)} \hfill& p_1  & 5,342  & 24,210
&\hfill 4.51\cr
 & \hbox{\bf \color{blue} Rest of economy} \hfill&  & &  &  \cr
9& \hbox{GDP (billion of 2015 RMB)} \hfill& G_2 & 4,768& 60,970 &
\hfill 12.79 \cr
10& \hbox{Labor force (billion)} \hfill& E_2 & 0.31  & 0.56 & \hfill 1.81\cr
\qtb
11 &\hbox{GDP per employee (2015 RMB)} \hfill& p_2 &
15,380&108,875&\hfill 7.07 \cr
\noalign{\hrule}
} $$
\vskip 1.5mm
\small
Notes: It can be observed that 
the rates of change of the labor force are an order of magnitude
smaller than the rates of change of the GDP. When intermediate
values are introduced one finds that both $ p_1 $ and $ p_2 $
increase as exponentials with exponents $ \alpha_1=0.062\pm 0.016 $  
(doubling time of 11.1 years) and 
$ \alpha_2=0.075\pm 0.011 $ (doubling time of 9.2 years). 
As a result, in the long-term the productivity gap 
increases as $ p_2-p_1 \sim \exp(\alpha_2t) $ 
\qL
{\it Sources: World Bank, Trading Economics, http://www.investopedia.com.}
\vskip 5mm
\hrule
\vskip 0.7mm
\hrule
\end{table}

The exponential increase of $ p_1 $ and
$ p_2 $ has two interesting implications.
\qbu As is well known, in a limited world
an exponential growth cannot last for ever which means that 
a saturation effect will set in sooner or later. The
data shown in Fig. 6 suggest that it should not occur before
2025.
\qbu Over past decades the
productivity gap between the agriculture sector and the rest
of the economy has increased exponentially:
$$ \hbox{gap}=p_2-p_1=p_2(1-p_1/p_2)=
p_2\left[1-(A_1/A_2)\exp(-d t)\right]\ d=\alpha_2-\alpha_1=0.013 $$
For large $ t $ the gap becomes equal to $ p_2 $
itself. Thus, any labor transfer from sector 1 to sector 2 will
trigger a productivity jump whose magnitude increases
exponentially in the course of time. 
This effect may compensate for diminishing transfers as $ e_1 $ 
becomes smaller. In other words, the productivity effect of
rural flight will remain significant even when $ e_1 $ will
fall under 10\% as may happen around 2030 according to Fig. 6.

\qA{Describing the GDP change due to a transfer of labor force}
Now, in order to describe the mechanism of Fig. 4 we must assume that
$ e_1 $ decreases by a quantity $ \Delta e_1 $. In fact, in what follows
it will be more convenient to follow the variable $ e_2 $
because its variations $ \Delta e_2=-\Delta e_1 $ are positive
in the course of time.\qL
The change
in $ G $ resulting from a transfer of population from sector 1 to
sector 2 will be:
 $$ G=G_1+G_2=p_1e_1E + p_2e_2E \rightarrow 
\Delta G=\left( p_1\Delta e_1 + p_2\Delta e_2\right)E
=(p_2-p_1)E\Delta e_2 \qn{1} $$

In the above expression of $ \Delta G $ the productivities
$ p_1, p_2 $ have been kept constant because we wish to
focus on the effect of a labor force transfer%
\qfoot{The huge changes which occur in the course of time for
$ p_1 $ and $ p_2 $ are a different question which will
be discussed later.}%
.
By introducing the factor $ k $, expression (1) becomes:
$$ \Delta G=
(k-1)p_1 E\Delta e_2 =(k-1)\left({ g_1\over e_1 }\right)G\Delta e_2
\rightarrow 
\left( { 1\over \Delta e_2 }\right){ \Delta G \over G }=
(k-1)\left({ g_1\over e_1 }\right) \qn{2} $$

We can take advantage of our previous observation regarding the stability
of $ g_1/e_1 $ to simplify expression (2) into:
$$ \left( { 1\over \Delta e_2 }\right){ \Delta G \over G }=
(k-1)<{ g_1\over e_1 }> \qn{3} $$

where $ <g_1/e_1> $ represents the average value
of $ g_1/e_1 $ over a period of several decades. 
For the United states, one would have $ <g_1/e_1> =0.69 $
whereas for China: $ <g_1/e_1> = 0.35 $.
\qpar
In the left-hand side the numerator and denominator can be
multiplied by 100 which allows to express both $ \Delta e_2 $ 
and $ \Delta G/G $ in percent. Thus the left-hand side represents
the percentage variation of $ \Delta G/G $ for a 1\% variation
of $ e_2 $.
\qpar

Note that equation (3) applies to any transfer of labor force.
In particular, it can also describe the {\it decline} in  GDP
that results from the transfer of labor force
from manufacturing to healthcare, a key-feature
in industrialized countries over the past three decades%
\qfoot{In this case sector 1 would be the whole economy
except healthcare and sector 2 would designate the healthcare
sector.}%
.
This point may be developed in a subsequent paper.

\qI{Labor force transfers as an engine of economic growth}

\qA{Rural flight seen in the broader view of all labor force transfers}

In the present paper we focus on the effect of rural flight
on economic growth for two main reasons.
\qbu This labor force transfer has a clear statistical
definition because of the fact that agriculture is a well defined
economic sector. In contrast, a sector such as healthcare
has a fairly fuzzy
statistical definition. For instance, do nursing homes belong
to healthcare or to the hospitality industry?
\qbu Historically rural flight has been of great importance
in all developed countries and nowadays it is still a major
factor in countries such as China and India.
\qpar

However, this focus should not hide the fact that rural flight
is only one case (albeit the most significant)
of a broad variety of labor force transfers which take place
continuously. For instance in developed countries two sectors
have had a rapid growth over the past 50 years: (i) the healthcare
and (ii) the information technology sector.
According to Table 1, around 2012
in the United States healthcare employed 
about 15\% of the total work force whereas the share of the
IT industry was of the order of one or two percent.
What impact on global productivity has had the development
of these sectors? Table 1 shows that healthcare's 
and IT's productivities were one half and twice the national
average respectively. 

\qA{A case in point: healthcare versus infrastructure development}

Across nations there are great disparities in the extension
of the healthcare sector. For countries with an aging populations
(as is the case of all developed countries) it is of course
natural to have an expanding healthcare sector but 
it must be recognized that its actual extent
is determined by a political decision.
Thus, in the United Kingdom which has a nationalized
medical system the weight of healthcare in the economy
is only one half of what it is in the US%
\qfoot{In spite of that, around 2009 life expectancy in the UK was
2.5 years longer than in the US.}%
.
\qpar

The fact that through its own poor productivity level, 
the healthcare sector drags down overall
productivity is not its only liability.  Because
most of its patients are elderly persons beyond
retirement age, health care has
little synergy with other activities.
\qpar

Completely different is the case of infrastructure
expenses. Many other sectors will benefit from
better transportation infrastructure. Clearly,
healthcare based development and infrastructure 
oriented development constitute two different
models. As one knows, the infrastructure model
is promoted by the Chinese government not only
domestically but also for the 
rest of the Asian continent. Regarding healthcare,
at this point it is too early to say
which system China will choose.
Will it be the US or the UK model%
\qfoot{One can be sure that this kind of decision will
be subject to lobbying efforts from many sides.}
?
Clearly this choice will affect the productivity
of the whole Chinese economy. Too much
resources and investments sunk in healthcare
will be a drag on the economy. 
At the beginning of the paper we said that one 
would expect a 8\% growth trend until 2025 but an
over extended healthcare industry 
may well slow down this expectation to 6\% or 5\%.

\qI{What fraction of growth does rural flight explain?}

In this section we compare the changes in GDP per employee
actually observed to those predicted by the
rural flight effect in three different ways. 
\qA{Correlation of relative variations}
Firstly, we wish to see if there is a connection
between the ups and downs in $ e_2 $ on the one hand and
the wage changes on the other hand. Obviously, 
cases (such as China)
in which there were little ups and downs will not be appropriate
for this test. In contrast, the test can be done
in good conditions for the case of the US because its
wage changes display several ups and downs. 
\qpar

The comparison between observed wage changes and those
predicted by the transfer model shows a significant
correlation (Fig. 5).
%
\begin{figure}[htb]
\centerline{\psfig{width=12cm,figure=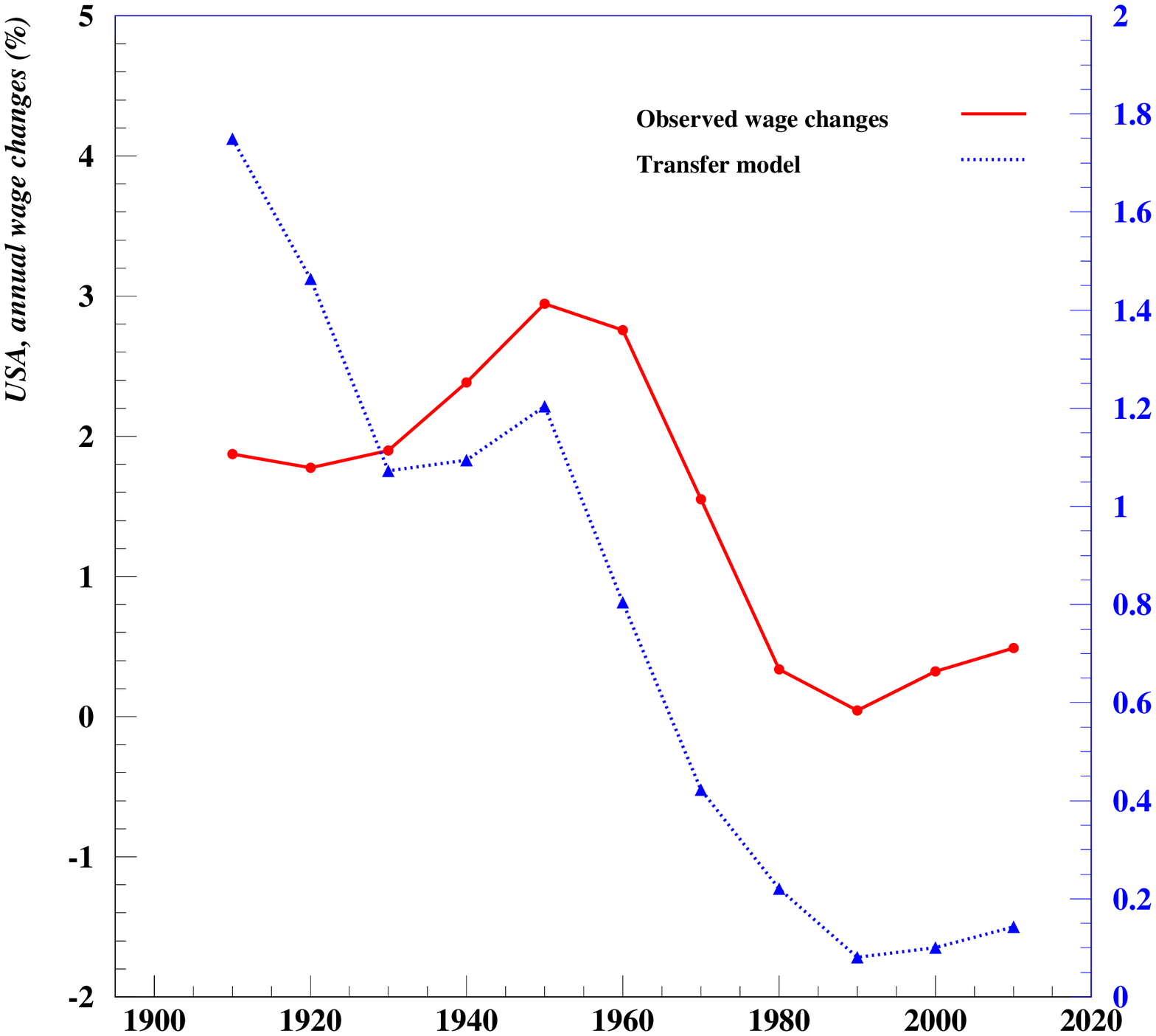}}
\qleg{Fig.\qhu 5\qhv Observed US wage changes versus
changes resulting from the transfer effect.}
{The correlation between the two series is
0.64 (95\% confidence interval is 0.08 to 0.90).
It can be observed that real wages did not fall
during the Great Depression. This is due to the fact that
the price consumer index fell faster than nominal wages.}
{Sources: Same sources as for Fig. 3.}
\end{figure}
%
Although one would like to
perform additional tests, it is
not easy to find appropriate cases. For instance,  
the income of
countries such as France, Germany or Japan which have
experienced wars on their territory will be affected
by exogenous shocks which have nothing to do
with the transfer effect.

\qA{Percentage of growth directly due to rural flight in 1990 and
  2015}

The correlation test performed in the previous subsection
was largely independent of the magnitude of the respective
changes. Now
we wish to determine what proportion of the increase in
GDP per employee 
can be accounted for by the rural flight effect.
For this test one needs a case in which the change rate
is as large as possible. China is well suited for this
purpose. Before making a formal calculation we will
compute the expected changes at the two ends of the time
interval 1990-2015. For this purpose we will use
the data given above (Table 3 and additional data).
\qpar

As already explained the rural flight effect is only one
of a number of transfer effects which contribute to
GDP growth. One of the clearest manifestations of the
effects of these transfers are the productivity
increases {\it within} the two sectors. This effect cannot be
explained by rural flight. Indeed, rural flight can only
explain the global growth of $ G $ due to the interaction
between agriculture and the rest of the economy.
The productivity
increases within the two sectors are due to interactions
between their various sub-sectors.
\qbu Tractors, trucks instead of horse driven carriages, 
engines, pumps, motorcycles,
mobile phones and many other items that are produced in
sector 2 progressively 
came into use in the country side, thereby lifting
$ p_1 $. These changes would happen independently of
any transfer of labor force. 
\qbu At the same time, in sector 2 outdated textile
factories were equipped with modern spinning and
weaving machines manufactured in other factories of sector 2,
newspaper printing machines were replaced by computerized
systems, supermarkets replaced small stores. As above,
these changes would happen independently of
any transfer of labor force.
Because sector 2
is more diversified than sector 1 it experiences
more synergy and that is probably
why the endogenous growth of sector 2 is faster than
the endogenous growth of sector 1.
\qpar

According to the comments following Table 3, we expect
a greater effect of rural flight in 2015 than in 1990.
Let us see if this is confirmed by observation.
\qpar

Over the whole period 1990-2015 the labor force in sector 1 
decreased by 170 millions which gives an average
change of 34 millions over every 5-year interval.
\qpar

{\bf \color{blue} 1990-1995} \quad Over this time interval,
one gets (everywhere RMB means ``RMB of 2015'')
\qbu The average of $ p_1 $ is 
$ \overline{p_1}=(5,231+7,434)/2=6,332 $ RMB 
\qbu  The average of $ p_2 $ is 
$ \overline{p_2}=(15,380+27,323)/2=21,351 $ RMB 
\qL
Thus, rural flight from 1 to 2 will increase $ G $
by: 
$ \Delta G' = 34\times 10^6(\overline{p_2}-\overline{p_1})=511\times
10^9 $ RMB.
\qpar
The actual increase in $ G $ was 
$ \Delta G=G(1995)-G(1990)= 5,627\times 10^9 $ RMB. Thus, the variation
due to rural flight represents: $ 511/5627=9.1\% $.
\qpar

{\bf \color{blue} 2010-2015} \quad A similar computation
leads to: $ \Delta G' =2,383\times 10^9 $ RMB, and
$ \Delta G=G(2015)-G(2010)= 5,627\times 10^9 $ RMB. In this case
the variation due to rural flight represents: $ 11.4\% $.
\qpar

As expected we see that in 2015 the change due to rural flight is a 
larger proportion of the total increase than in 1990. We also see
that the largest part of GDP growth must be attributed
to endogenous increases of $ G_1 $ and $ G_2 $ which in turn are
due to transfers occurring {\it within} these sectors.

\qA{General formulation}

In this subsection the previous computation is presented
in a broader and more formal way.
\qpar
Let us denote by $ p $ the GDP adjusted for 
inflation and per employee. From 1985 to 2014, $ p $
has increased regularly at an average annual rate of
9.2\%; for the whole interval
this rate resulted in a multiplication by 10.3
\qpar

The productivity multiplier $ k =p_2/p_1 $ fluctuated between
4 and 6. 
In order to get an upper bound of the growth
accounted for by rural flight, we will
give $ k $ its maximum value of 6. As a first step,
consider the two extreme situations: $ e_1=1 $
and $ e_1=0 $. Moving from the first to the second would
result in a multiplication of $ G $ by 6. In other words,
in this thought experiment 
the transfer effect explains only $ 6/10.3=58\% $ of the growth.
\qpar 

In order to get a more realistic estimate we write
the values of $ G $ in 1985 and 2014 in the following way
($ G' $ denotes the values of $ G $ predicted by rural flight):
 $$ G'(i)=p_1(i)E_1+6p_1(i)E_2=\left[e_1(i)+6e_2(i)\right]p_1(i)E $$

With $ e_1(1)=67\% $ and $ e_1(2)=31\% $ one gets:
 $ G'(2)/G'(1)=1.67\left( p_1(2)/p_1(1) \right) $.
In the rural flight model the productivity $ p_1 $ is supposed fixed,
that is $ p_1(1)=p_1(2) $. Under this assumption
the transfer of labor accounts for only
$ 1.67/11=15\% $ of the actual productivity growth.
\qpar

Equation (3) would lead to a similar conclusion.
For average annual changes it gives:\qL
$ \Delta G'/G'=(k-1)<g_1/e_1>\Delta e_2=5\times 0.4\times 1.2=2.4\% $\qL
whereas the actual annual GDP growth is $ \Delta G/G=9.2\% $.

\qA{Implications}

The fact that rural flight accounts for only a fraction of the
total growth 
has two important implications.
\qbu A recession may temporarily stop employment shifts
from sector 1 to sector 2. If this transfer would account for
a high percentage (say some 80\% or more)
of economic growth, then its discontinuation
would freeze further growth altogether 
and transform the recession into a depression.
However, if employment transfer actually is 
only a secondary factor
the synergy engine will continue to work and will pull
the economy out of the recession%
\qfoot{Needless to say, economic growth can be thwarted
for many other reasons, e.g. deflation, capital flight,
disruption of the banking sector and so on.}%
. 
\qL
This is indeed what was
observed in developed countries in the decades following World 
War II. There were only mild recessions.
In other words, unless there is a political 
upheaval no major interruption of
growth should be expected in China for
at least 15 years, that is to say until it has become a service
economy too.
\qbu As seen earlier, in developed countries
the growth of wages stopped or slowed down in the 1980s.
This was due to the shift of their economies toward
a service sector economy which is dominated by low productivity
sectors of which healthcare is an important component.
The fact that rural flight almost came to an end in the 1980s
was due to two circumstances: (i) The labor force of sector 1 
was already largely depleted.
(ii) Because productivity growth in sector 2 was slowed down,
this sector lost its attractivity in terms of higher
salaries. Indeed, the growth of sector 2 was largely due to
the introduction of so-called ``odd jobs'',
``second class jobs'' or ``deskilled casual work''.

\qI{Conclusion}

\qA{Chinese economic growth over 2015-2035 in comparative perspective}

Several interesting features are highlighted in Fig. 6.
\qbu  The 
transfer of labor force from agriculture to the industrial
and service sectors has progressed unabated in all
developed countries. Economic crises or even the two world wars
may have slightly slowed down the evolution (as in the
case of Austria for instance) but they did not have 
any lasting impact.
%
\begin{figure}[htb]
\centerline{\psfig{width=12cm,figure=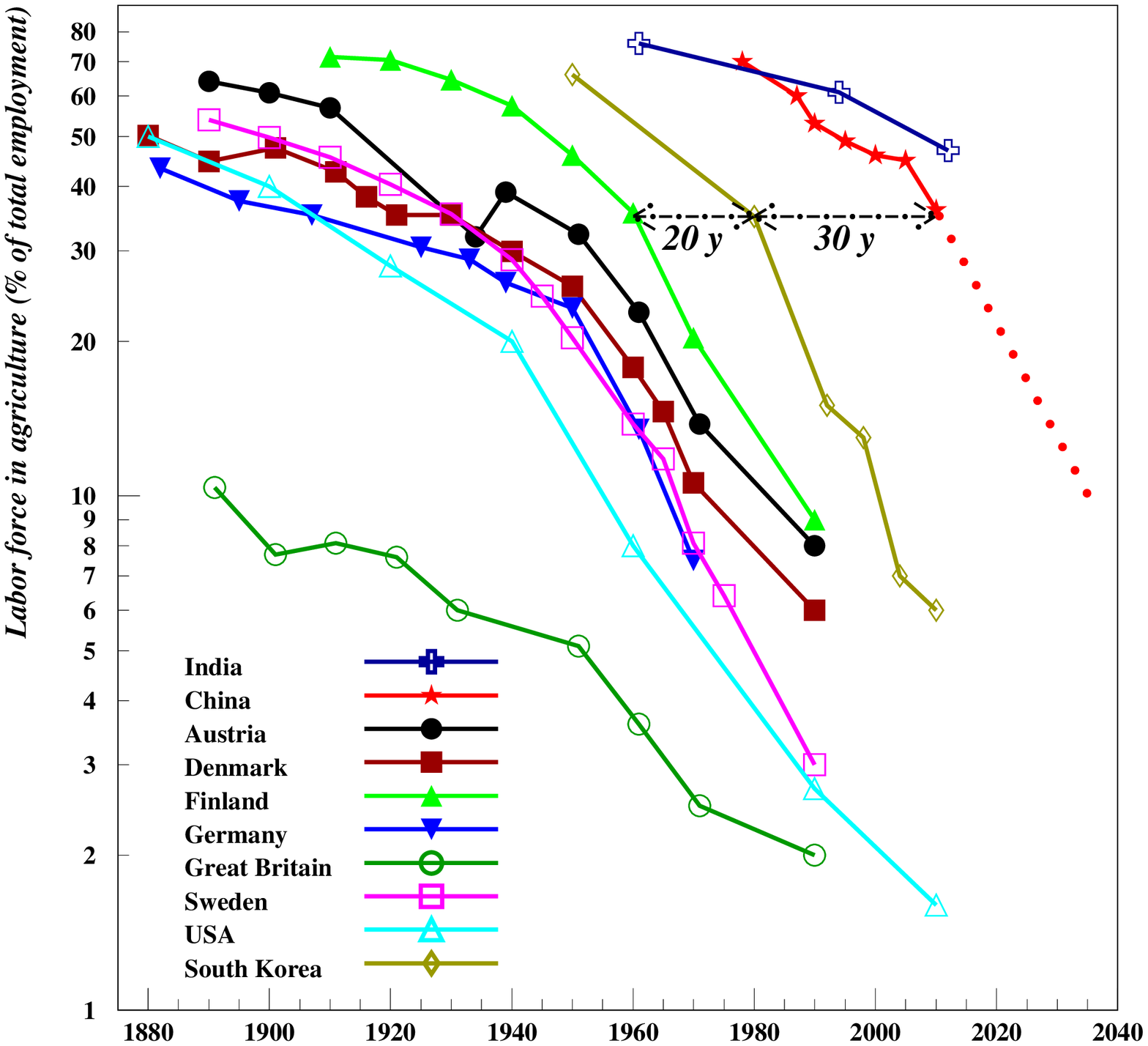}}
\qleg{Fig.\qhu 6\qhv Labor force in agriculture.}
{Apart from the UK which stands apart, all developed
countries experienced more or less the same evolution.
The fact that the curves for Sweden and Germany are
almost the same shows that wars had almost no
impact. The curve for China was extrapolated based on the
rates seen in other countries and particularly in
South Korea. Regarding the special case of Britain see the
explanation given above in relation with Fig. 3.}
{Sources: Until 1970: Flora et al, (1987) and Historical
Statistics of the United States (1975).
After 1970: Trading Economics website and Saint Louis
Federal Reserve.}
\end{figure}
%
\qbu We have already suggested
that South Korea may be a good predictor of the future
of the Chinese economy. 
The development of South Korea followed the pattern
set by developed countries. 
We see that in terms of proportion of the labor force working
in the agricultural sector
Finland was about 20 years ahead of South Korea
which itself was 30 years ahead of China%
\qfoot{In this respect one should not forget that, as in Taiwan,
the Japanese occupation had resulted in a substantial
industrialization not only in the north of Korea but also
in the south; more evidence can be found in Roehner (2015).}%
.
If one uses the rate of South Korea to extrapolate the curve 
for China one arrives to the conclusion that in 2035
the percentage of the labor force in agriculture will be
about 10\%.
\qbu Among developed countries the UK is a case apart.
Its agricultural sector
fell to a level of 10\% some 60 years before the United
States.
In this sense, Britain was the first
post-industrial economy; this may explain the fact
(observed in Fig. 3) that its rate of wage increase was
substantially lower than in other countries.
\qpar

What was the evolution of agricultural employment in the USSR
and later on in the Russian Federation? By collecting 
data from various sources (Gaidar 2012, W\"adekin 1982,
Lerman et al. 2003, Trading Economics website) 
we got the following picture:

\begin{table}[htb]
\small
\centerline{\bf Table  4\quad Russia: Employment in agriculture
in percentage of total labor force}

\vskip 3mm
\hrule
\vskip 0.7mm
\hrule
\vskip 2mm
 $$ \matrix{
\qtb
 1930 & 1960 & 1978 & 1985 & 1990 & 2000 & 2010 & 2014 \cr
\noalign{\hrule}
\qth \qtb
87\% & 39\% & 21\% & 15\% & 14\% & 15\% & 9\% & 7\% \cr
\noalign{\hrule}
} $$
\vskip 2mm
\hrule
\vskip 0.7mm
\hrule
\end{table}

The table shows that the interruption of the fall
of the share of agricultural employment coincided with the
crisis of 1988-2000. So, once again, we see that
this variable
is a good indicator of long-term economic growth.
We see also that the time interval 1930-1960 saw a massive
reduction in agricultural employment and we know that 
it was accompanied by rapid economic development.

\qA{Rural flight seen as a good indicator of productivity growth}

In the first part of the paper
it was shown that a wage transition occurred in most
developed countries in the 1980s. After decades of rapid growth,
wages started to slump.  Rural flight appeared as
a good candidate for
explaining not only this wage transition but also the rapid
growth observed in China in past decades.
Incidentally,
one should not think that there was no or little
growth before 1983. According to Chinese GDP statistics,
the real average annual GDP growth rate was $ 9.8\% $ from 1952 to
1982 and $ 9.9\% $ from 1983 to 2014. Even though
national accounting methods may have been different 
before 1983,
the idea of a stagnant economy before 1983
seems inappropriate.
Actually, the most conspicuous difference
between the two periods is that in the first one growth was
much more irregular than in the second. The coefficients
of variation of the annual growth rates were 194\%
and 27\% respectively. 
\qpar
In order to test the  idea of a connection between rural flight
and economic growth
we set up a model which allowed us to predict the 
change of income resulting from rural flight.
It was seen that this model is in
good qualitative agreement with
observation. As expected, rural flight accounts for only
a fraction of total growth but it is also a good indicator
of the dynamism of sector 2. If tomorrow the
Chinese economic growth starts to create a lot of ``second class
jobs'' (for instance by an over-extension of the healthcare
sector) then, those young people who presently still remain in
Chinese villages (see Appendix C) will certainly think 
about it twice
before moving to cities. This, in turn, will make economic growth
more sluggish. So, the main issue is whether 
in the coming decades China will be able to
avoid the ``odd job trap''.

\qA{Wage progression and labor inflow}

It may seem surprising that in this paper we did not
consider the question of immigration.
As for any other good, the cost of labor is affected by supply and
demand changes. Supply is controlled by labor force expansion
either endogenously or exogenously (i.e. through immigration)
whereas demand is conditioned by GDP growth. This
question would deserve a separate study but here we will limit
ourselves to a few observations.
\qbu Due to the sheer size
of the Chinese population foreign  immigration
will have little incidence in China. 
The only country whose population
could match the population of China is India. However, whereas
there are important Indian
communities in the UK and US, immigrants to China
would have to learn Chinese, quite a formidable
challenge for them.
\qbu The case of Japan suggests
that at the level of a whole country (as
opposed to a specific sector) labor inflow through immigration
is rather a second-order effect. Indeed, 
despite a much lower immigration rate
Japanese wages leveled off in the same way as in European countries
(see Fig. 3). 
\qbu In contrast, at sectorial level, immigration
may play a role. In this respect, as a case study,
it would be interesting to
investigate the impact on labor cost
of the inflow of Filipino nurses into the US%
\qfoot{Their immigration
is facilitated by the fact that their curriculum
integrates the requirements and specifications of the US healthcare
industry which includes practising the English
language.}%
.

\vskip 4mm
{\bf Acknowledgments}\quad We wish to express our gratitude
to Ms. Corina Neuerer of the ``Statistisches Bundesamt''
(German Federal Statistical Office) and to Ms. Alyson Williams
of the British ``Office of National Statistics'' for 
helping us to locate and retrieve some of the statistical series
used in this paper.

\appendix

\qI{Appendix A. Metrics for measuring income}

Income can be measured by various indicators
and as the findings reported in
this paper are, at least to some extent, indicator-dependent 
it is important to keep in mind the differences between
them. They are summarized in the table below.

\begin{table}[htb]

\small

\centerline{\bf Table A1 \quad Different metrics for measuring 
personal income}

\vskip 5mm
\hrule
\vskip 0.7mm
\hrule
\vskip 2mm

$$ \matrix{
\qtb
&\hbox{Indicator}\hfill &  \hbox{\it Comment}\hfill \cr
\noalign{\hrule}
\qth
1 &\hbox{GDP/capita}\hfill & \hbox{\it Apart from income
in the form of salary, the GDP also involves capital gains,}\hfill \cr
 & \hbox{}\hfill & \hbox{\it such as dividends or profits.
In recent decades this part of national income has} \hfill \cr
& \hbox{}\hfill & \hbox{\it been growing. Today in the US non-salary
income represents around 50\% } \hfill \cr
2 &\hbox{GDP/employee}\hfill & 
\hbox{\it Corrects the fact that GDP/capita increases when}\hfill \cr
 &\hbox{}\hfill & 
\hbox{\it housewives take up a full time job.}\hfill \cr
3 &\hbox{Average earnings}\hfill & 
\hbox{\it In contrast with the median, the average is highly sensitive
to top earnings}\hfill \cr
4 &\hbox{Median earnings}\hfill & \hbox{\it }\hfill \cr
5 &\hbox{Average wage}\hfill &  
\hbox{\it The terms ``worker'' or ``manual worker'' 
usually mean non-managerial employee;}\hfill \cr
 & \hbox{of workers} \hfill & \hbox{\it compared to services,
manufacturing provides better protection to workers} \hfill \cr
 & \hbox{in manufacturing} \hfill & 
\hbox{\it particularly because of remaining unionization.} \hfill \cr
6 &\hbox{Average wage of}\hfill &  
\hbox{\it Includes (non-managerial) service employees.}\hfill \cr
\qtb
 &\hbox{all workers}\hfill &  
\hbox{\it }\hfill \cr
\noalign{\hrule}
} $$
\vskip 1.5mm
\small
Notes: Broadly speaking these different incomes
decrease from top to bottom. Average earning is similar to 
the macroeconomic variable
called ``personal income'' (i.e. basically GDP minus capital depreciation
minus corporate profit divided by employment).
At the bottom of the table
an additional distinction may be appropriate 
between hourly, weekly and 
monthly wages especially in case of large changes in weekly hours 
worked.
\vskip 5mm
\hrule
\vskip 0.7mm
\hrule
\end{table}

\begin{figure}[htb]
\centerline{\psfig{width=15cm,figure=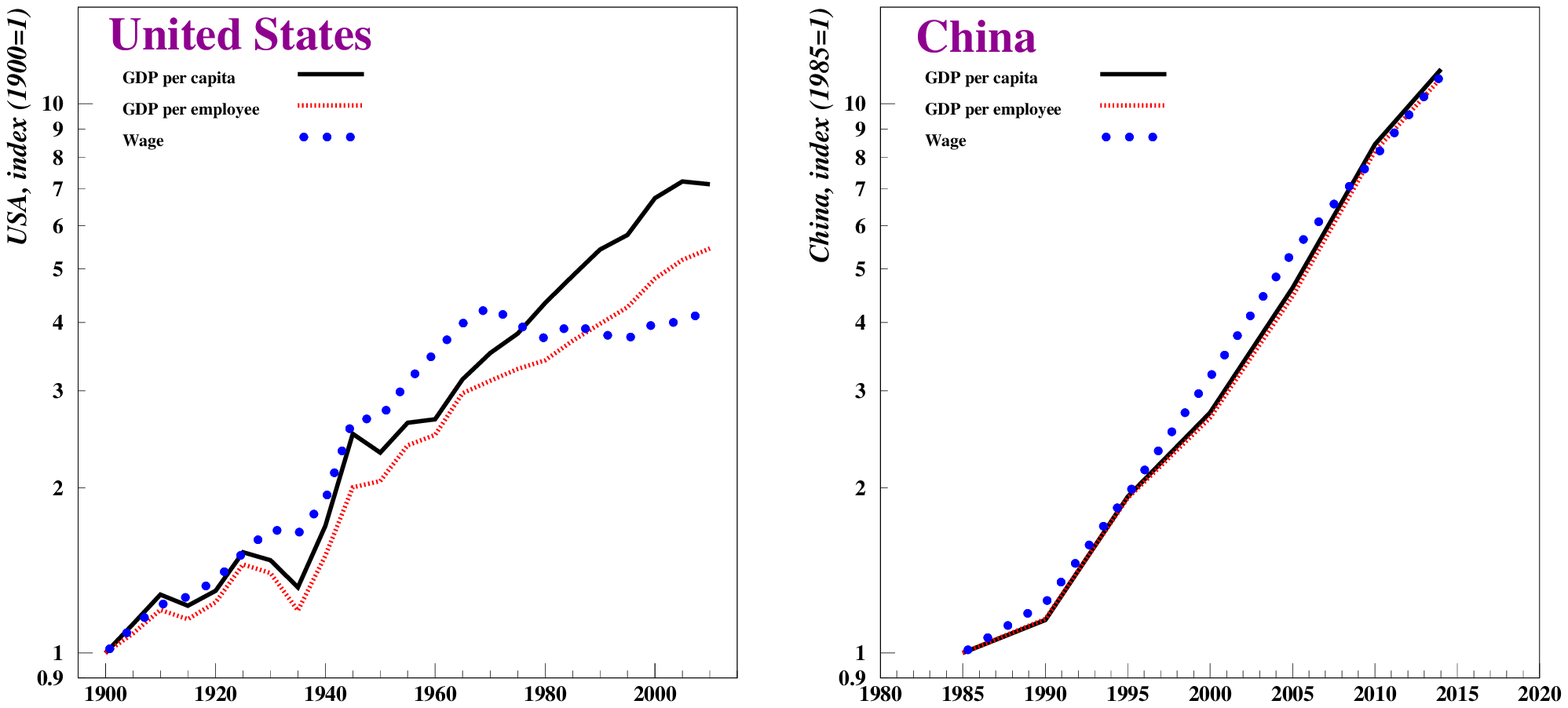}}
\qleg{Fig.\qhu A1\qhv Comparison of different metrics
measuring income for the United States and China.}
{All incomes are expressed in national currency and
adjusted for inflation.
The two countries are not at the same stage in terms
of overseas investment and financialization
of the economy; therefore,
the fact that so far in China there has been no
divergence between GDP/capita and wage level
does not mean that it will be the same in the future.}
{Sources: USA: Historical Statistics of the United States (p.224),
Saint Louis Federal Reserve website, Liesner (p.98-99),
TradingEconomics website; China: http://www.chinability.com/GDP.htm,
TradingEconomics website.}
\end{figure}

\begin{figure}[htb]
\centerline{\psfig{width=15cm,figure=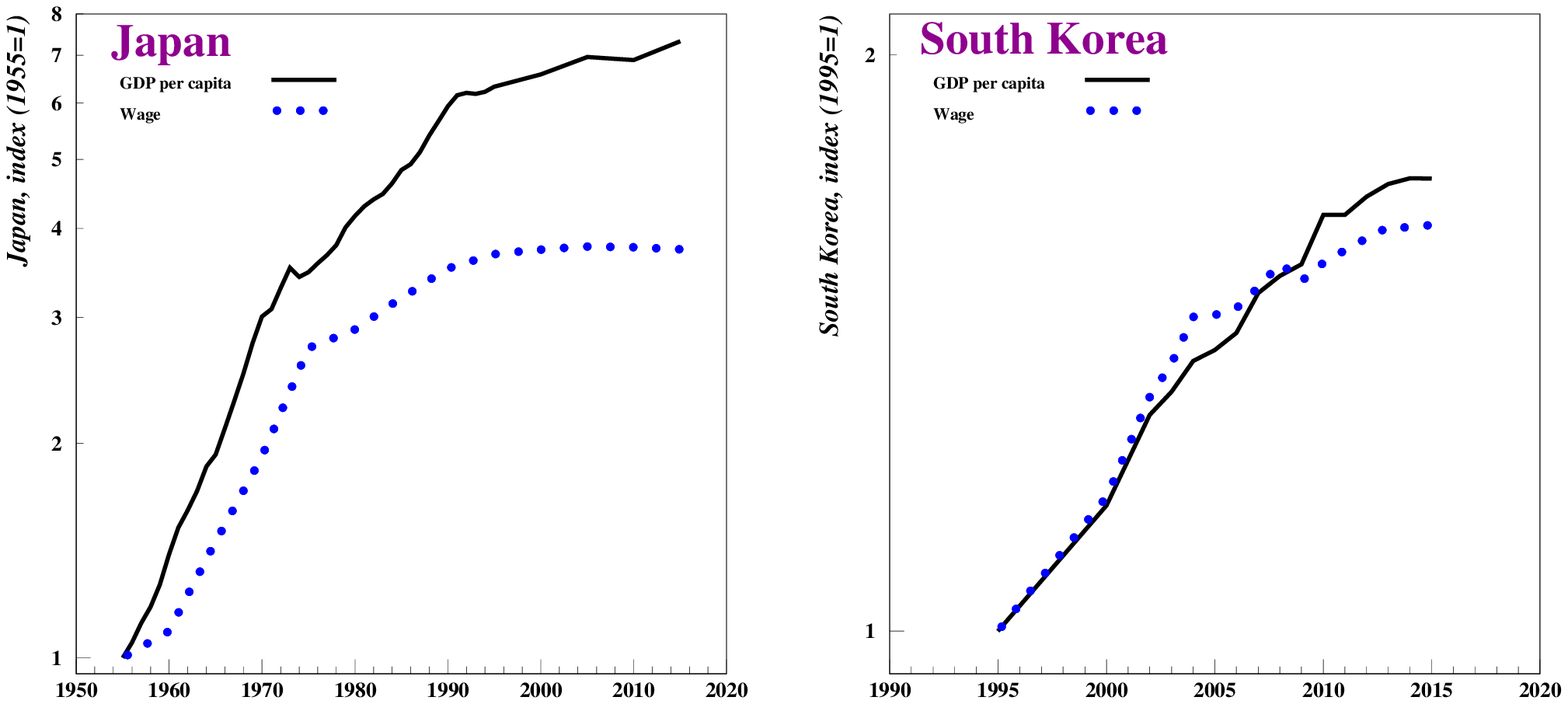}}
\qleg{Fig.\qhu A2\qhv Comparison of different metrics
measuring income for Japan and South Korea.}
{All incomes are expressed in national currency and
adjusted for inflation.
The curves of GDP per employed person were omitted
because labor force as a percentage of population showed
no definite trend. Thus, in South Korea it fluctuated
randomly between 58\% and 60\%. The big gap observed
in Japan between GDP per capita and wage came as a surprise.}
{Sources: Japan: Historical Statistics of Japan (website),
Trading Economics website (GDP at constant prices in
billion of yen); South Korea: Korean Statistical 
Information Service (KOSIS), statinfo.biz. It would
be useful to find wage data series starting before 1995.}
\end{figure}

For a better understanding of the difference between GDP/capita
and wages it is useful to remember that the GDP has
basically 3 components: (i) compensation of employees, (ii) operating
surplus (i.e. mainly profits) and (iii) consumption of fixed
capital (i.e. depreciation of equipment). Altogether these
components usually account for 90\% of the GDP; the first component
is of the order of 50\% and in most countries it
has been decreasing over past decades.
\qpar

As an illustration of the difference between median and average income,
one can mention that in Singapore
in 2014 the average monthly household income was $ 9,176 $ Singapore
dollars, whereas the median was only 6,500, i.e. some $ 30\% $ less.
In addition, it should be noted that most income statistics
published by the Singapore statistics office include the ``employer
CPF contribution''. The acronym CPF means ``Central Provident Fund''
and it designates a social security fund destined to cover the
retirement, healthcare, and housing needs of the employees.
Currently, it represents about 11\% of the wages. 
A similar system exists in many countries.
It can be noted that as
health care expenses are much higher for elderly employees
it means that they benefit more from CPF transfers
than younger employees.

\qI{Appendix B. Data sources}

As for any comparative research, one of the main challenges of this
work was to find statistical data pertaining to different countries 
but recorded in such a way as to allow meaningful comparisons.
Fortunately the Internet
provides access to the websites of the
statistical offices of many countries and of
international organizations such as the Work Bank or the OECD.
However, data comparability often remains a tricky issue.
\qpar

Basically, for our study we needed data for the following variables.
\qbu Average wages.
\qbu Price consumer index (CPI) in order to deflate 
the nominal wage data into a series of real wages.
\qbu Distribution of employment by industry.
\qpar

Comparisons are hampered by several difficulties.
\qee{1} Depending upon the country,
the average wage may be defined in different ways:
hourly wages (US), weekly wages (England and
Germany), annual wages (China). In order to convert
hourly wages into annual wages one must know the average weekly
working hours. 
\qee{2} Much more problematic is the question of who are the
recipients of the corresponding wages. The problem can
be illustrated by the case of Germany for which one has fairly
detailed information. The ``Federal Statistical Office''
publishes four quarterly series of wages. 
(i) The first gives weekly nominal hourly wages paid to manual workers
in manufacturing. (ii) The second gives real wages 
paid to the same workers. (iii) The third gives real 
hourly wages paid to all employees. (iv) The fourth gives
real monthly wages paid to all employees. Between 1991 and 2011,
the first increased by 74\%, the second by 22\%, 
the third by 11\%, and the fourth by 1\% 
(International Labour Organization 2014, p. 47).
Even, within the same series there can be changes in the course
of time. Thus, the first series cited above referred to manual workers
between 1980 and 2006 but to skilled workers after 2007.
\qpar
Such observations lead to the conclusion that one should not expect
an accuracy better than $\pm 10\% $ and that one should 
adopt a statistical methodology in line with this 
level of uncertainty. 
For the investigation of broad long-term
trends as conducted in this paper the poor comparability of the data 
is not a major obstacle.

\qI{Appendix C. Workforce reserve in rural China}

\qA{Defining the question}

The mechanism on which this paper is based consists in
transfer of population
from rural to urban areas. It was observed that
in industrialized countries productivity was boosted
through this transfer and that this effect was significant 
as long as the rural population 
represented a sizable proportion of total population,
basically over 5\%. This argument assumes implicitly
that the rural population comprises persons who can
take up a job in the manufacturing or service sector once
they have moved to cities. There are two situations
in which such an assumption does not hold.
\qee{a}  The first case is
when the rural population comprises only elderly
persons beyond retirement age (Fig. C1b).
This situation would
occur when all middle-aged persons have moved to cities
together with their children. In this case there would
be no productivity reserve whatsoever in spite of
a rural population which may be well over the 5\% threshold.
\qee{b} The second case is
when the rural population comprises only
grand parents and their grand children (Fig. C1a).
This situation would
occur when all middle-aged persons have 
moved to cities but have left their
children behind them in their villages%
\qfoot{In societies in which house wives are not
supposed to take up a job, one could assume that
they remained in their villages with their children. 
In this case 
there would be no productivity reserve despite
a substantial rural population.}%
.
In this case, there is no short-term productivity reserve
but over a period of one decade the young segment of the
rural population will contribute to productivity progress.
\qpar

In former rural flights (described in Fig. 6
for industrialized countries) the situations (a) and (b) 
occurred only marginally and, as a result, productivity
progress was not substantially slowed down. 
Why should
the picture be different in China? 
\qpar
One obvious reason can be mentioned
which is the pace of the rural flight in China.
The very same evolution that took one century and a 
half in western countries
is taking place in China in less than 50 years.
So, in order to make sure that the rural flight
mechanism will have the effect that we expect we must
consider more closely the changes in the composition
of the rural population.
We will do that in two steps: first we give a fairly
qualitative picture, then in a second step we 
show two graphs which illustrate
age-specific composition changes.

\qA{Qualitative picture}

This subsection is based on a paper in 
Chinese by Zong and Xiang (2013).
As this paper provides many interesting tables and graphs
one may wonder why we use it here 
to give a {\it qualitative} picture. 
It is because the paper is not based on census data
but rather on a survey of their neighbors 
carried out by Chinese graduate students when returning
to their home place during their term break%
\qfoot{At the geographical level it is limited to 
5 provinces of central China, namely: Anhui, Henan, Hunan,
Jiangsu and Sichuan. All results given below are
averages over these provinces.}%
. 
As a result,
the data are based on fairly small samples of one or
two thousands.
However, they provide a much more detailed view
than could be obtained from census data. The following
points are of particular interest.
\qbu Instead of a simplistic binary picture the paper
considers several cases: persons who
moved away and are no longer registered in the village ($ A $);
persons who are still registered but are present in the
village during less than 3 months per year ($ B_1 $); this
category would correspond to students or young people who
return to their village a few times every year.
Three similar categories ($ B_2,B_3, B_4) $ correspond to 
persons who stay in the village more than 3 months but 
less than 6 months, more than 6 months but less than 10 months,
and more than 10 months. Around 2010 the percentage
distribution was as follows:
$$ A:\ 29\%,\quad  B_1:\ 18\%,\quad  B_2:\ 3.2\%,\quad  
B_3:\ 2.7\%,\quad B_4:\ 47\% $$
Not surprisingly, the age group which had the smallest proportion
of $ B_3+B_4 $ was $ 20-24 $; however, even in this age group
the $ B_3+B_4 $ represented about 30\% of the whole $ B_1,\ldots B_4 $
group.
\qbu If one considers children under 15, about 50\% lived
with their two parents, 22\% lived only with their mother,
and 25\% lived with their grandparents%
\qfoot{A negligible proportion of 1.7\% lived only with their
father.}%
.

\qA{Quantitative picture}

The age distributions shown in Fig. C1a,b are based on the 
censuses of 1982, 2000 and 2010. They confirm the picture 
suggested by the previous subsection in the sense that 
the demographic situation in rural areas is still fairly remote
from the idealized cases represented by the dotted curves.

\begin{figure}[htb]
\centerline{\psfig{width=16cm,figure=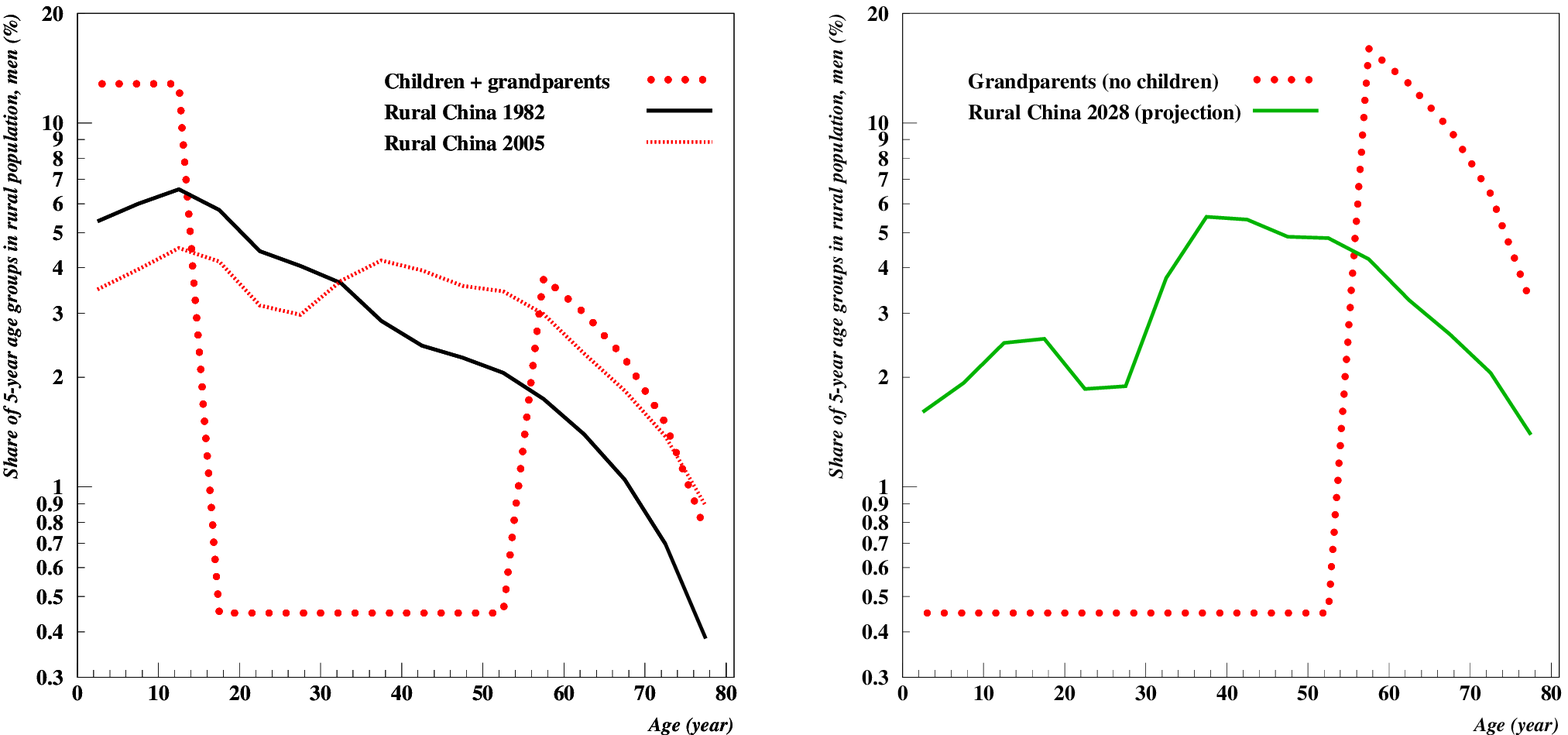}}
\qleg{Fig.\qhu C1a,b\qhv Age distribution in rural China.}
{The graphs compare two idealized situations (dotted curves)
with the actual age distributions. The expression ``grandparents''
designates persons over the age of 55 that is to say persons
who can no longer be counted in the manpower reserve 
susceptible of moving to cities.
The curve for 2028 is
a linear projection based on the data of 1982 and 2005. It can be seen
that although closer to the ``grandparents only''
picture, it is still  fairly remote from it. Note that because of
the $ y $ log scale the flat parts
of the dotted curves that should be at zero percent had to be set
to a low but non-zero percentage.}
{Sources: Kasperkevic (2012), Zhong (2013), Yiliminuer (2015).}
\end{figure}

Can we make comparisons with other countries?
In the United States around 1930 the rural population
represented 44\% of the whole population. In the rural
population the share of the group under 15 was 35\%.
For China around 2010
the corresponding figures were 55\% and
13\%. The first number confirms what we already emphasized
namely that rural flight is only at its beginning in China.
Comparing the shares of the age-groups under 15 makes
little sense because it simply shows that in the US
the birth rate and immigration rate were much
higher than presently in China.
\qpar

In conclusion, we can say that there is still a substantial
manpower reserve in rural areas that will become available
for moving to cities as soon as higher agricultural
productivity will make that possible.

\qI{Appendix D. From personal to social productivity}

At the end of the paper it was observed that progress in
productivity comes mainly from 
what we called a synergetic effect. In the present section
we try to explain this effect in more detail, 
at least in a qualitative way.

\qA{Productivity at individual level}

The simplest measure of the productivity of labor is through the ratio
of GDP to the total number of hours worked. According to this
definition it is fairly clear that the transfer of an agricultural worker 
to a modern factory will result in his productivity being 
substantially improved. This was true in the 19th century, when
Irish tenants came to work in Birmingham, Liverpool, or Manchester.
This same mechanism operates when a Chinese worker moves
from his (or her) village to a modern factory in Guangzhou or
Shanghai, but in this case the effect is much stronger for indeed
the productivity of a 21st century textile factory will be way
higher than the productivity of a textile factory of the 19th 
century.
\qpar

As demonstrated by Adam Smith's needle factory example,
higher productivity can be obtained by decomposing
the production process into simple steps that can then be
standardized and mechanized. At the other end of the spectrum
are tasks which so far have not been subjected to the same process.
A classical example given
by the French economist Jean Fourasti\'e is the ``production''
of an haircut.
Fourasti\'e observes that the time required for
an haircut, namely about 20 minutes, did hardly change over the
past two centuries. It is important
to observe that the reason of this
stagnation is more social than technical. It would certainly be
possible to build a machine able to perform a haircut in a few 
minutes. However, such machines would perhaps not be well
accepted by customers, nor would they be welcomed by haircutters.
Similarly it would be fairly easy to set up online
visits to the doctor, but it can be expected that such a move
will not be welcomed by the medical profession because 
doctors in developed countries would have to compete
with doctors from developing countries. 
\qpar
We mentioned the previous examples in order to emphasize that
the improvement of productivity is not only a technical
challenge but that it has also an important social dimension.
This leads us directly to the notion of social productivity.

\qA{Social productivity}
If we consider productivity not only at the level
of a company or an industry but at the level of the 
whole society
the picture changes completely.
For the sake of the present discussion we assume that 
a capital $ C $ (say one million euros) is spent in different ways.
Let us examine the possible effects on the productivity of
the whole country.
\qbu {\color{blue} Type -1: Frozen capital, negative effect.} \quad
In this first operation
the capital $ C $ is kept at home by its owner and not spent in
any way. Thus, $ C $ is simply withdrawn from the 
money in circulation. This reduction of the money supply will
have the usual deflationary effect. From an economic perspective
this operation is the most negative way of using (or rather not
using) the capital. The same negative effect would be achieved
if the capital is sent abroad without any return.
\qbu {\color{blue} Type 0: Capital re-injected into the economy but with
no positive benefit.}\quad The classical example of such an 
operation is the broken window fallacy introduced
by the French economist Fr\'ed\'eric Bastiat in 1850. 
Bastiat considers a shopkeeper whose careless son happened to
break a pane of glass and he describes bystanders who say
``Everybody must live, isn't it?
What would become of the glaziers if panes of
glass were never broken?''. Bastiat comments that by paying the
glazier the shopkeeper puts the money in circulation, but at the
same time he emphasize that had the window not be broken, then
this expense could have been
affected to an usage of greater social usefulness for the
{\it society as a whole}. This brings us to the case 
of an expense which has a positive effect on the society.
\qbu {\color{blue} Type 1: Capital outlay which 
has a positive effect on social productivity.}\quad
Suppose that the owner of a firm spends the capital $ C $ on
buying several new trucks which will allow easier delivery of
the goods produced by the firm. Not only will the 
capital be re-injected into the economy but 
throughout the trucks' life-time, 
it will allow smoother and faster delivery. In short, 
the production and usage of the trucks will
improve the productivity of the firm and of society as a whole.
The same observation applies to other investments
which will help doing things faster and better. 
As an illustration at household level, one can mention the
fact that
the production and usage of washing machines
will save much time compared to hand washing.
In contrast, using bigger plasma TV screens will have
little effect on social productivity; actually it may have
a negative effect in the sense that children will spend more time
watching TV at the expense of their school work.
The same observation may apply to other innovations of present
time, e.g. video games or mobile phones.
\qpar

In a general way,
it is certainly true
that most devices produced by industry will at the same time
enhance the productivity of the society. This is of course true
for capital goods such as machine tools or tractors
but also for
many products bought by the public, 
e.g. clocks, computers, cars
and so on. In these cases, there is a kind of synergy
through which the technical and scientific capabilities
of a population increase.
On the contrary, most services provided
by the tertiary sector are of type 0.
It is true that many of them
(e.g. healthcare) cannot be avoided and that others such as
cultural services are enjoyable, but this does not remedy
their weak (or sometimes negative) contribution to global 
productivity.
\qpar
An additional word is in order regarding the computer and
telecommunication industry. Whereas the production of computers
and other telecommunication devices is naturally included in 
the manufacturing sector it is not always clear how the
software industry is categorized. On account of the fact that
software applications are sold along with computers,
digital machine tools or 
mobile phones, it would be logical to include the software 
industry into the manufacturing sector. 
The synergy which links together software applications 
is another reason for including the software industry into
the secondary sector rather than into the service sector.

\vskip 5mm
{\bf References}

\qparr
Bajpai (P.) 2014: China's GDP examined: a service sector surge.\qL
http://www.investopedia.com (31 October 2014)

\qparr
Bastiat (F.) 1850: 
Ce qu'on voit et ce qu'on ne voit pas, ou, 
l'\'economie politique en une le\c{c}on.
Guillaumin, Paris. English translation:
``Things seen and things not seen''. Cassel and Company, 1904.

\qparr
Flora (P.), Kraus (F.), Pfenning (W.) 1987: State, economy, and
society in western Europe, 1815-1975, volume 2.
Macmillan, London.

\qparr
Fourasti\'e (J.) 1951: Machinisme et bien-\^etre. Editions de
Minuit, Paris. Translated into English under the title:
``The causes of wealth'' (1961). 
The Free Press, Glencoe (Illinois).

\qparr
Gaidar (Y.) 2012: Russia, a long view. MIT Press, Cambridge 
(Massachusetts)

\qparr
International Labour Organization (ILO) 2014: Global wage report 2012/2013.
Wages and equitable growth. International Labour Office. Geneva.

\qparr
Kasperkevic (J.) 2012: The demographic crisis is going to
hurt most in rural China. 
Business Insider (17 April 2012).

\qparr
Lerman (Z.), Kislev (Y.), Biton (D.), Kriss (A.) 2003:
Agricultural output and productivity in the former Soviet Republics.
Economic Development and Cultural Change 51,4,999-1018.

\qparr
Liesner (T.) 1989: One hundred years of economic statistics. Facts
on File, New York.

\qparr
Roehner (B.M.) 2009: Hidden collective factors in speculative
trading. Second edition. Springer-Verlag, Berlin.

\qparr
Roehner (B.M.) 2015: Relations between US forces and the population
of South Korea. Working report LPTHE, University of Paris (UPMC).

\qparr
Shanmugaratnam (T.) 2015: Budget speech given in Parliament.
The Straits Times 23 February 2015, article by Chew Hui Min.

\qparr
W\"adekin (K.-E.) 1982: Agrarian policies in Communist Europe.
A critical introduction. Martinus Nijhoff, The Hague.

\qparr
Yiliminuer (T.) 2015: La Chine, championne du monde de la
g\'erontocroissance? Population et Avenir 1,14-16.

\qparr
Zhong (F.-N.), Xiang (J.) 2013: A regional comparison of the
age distribution in the rural Chinese population and its
implications to policy making [paper in Chinese, original
title is: Woguo nongcun renkou nianling jiegou de diqu bijiao
ji zhence hanyi].
Modern Economics Research 3,5-10.

\end{document}